\documentclass{article}
\usepackage{ijcai26}

\usepackage{times}
\usepackage{soul}
\usepackage{url}
\usepackage[hidelinks]{hyperref}
\usepackage[utf8]{inputenc}
\usepackage[small]{caption}
\usepackage{graphicx}
\usepackage{amsmath}
\usepackage{amsthm}
\usepackage{booktabs}
\usepackage{algorithm}
\usepackage{algorithmic}
\usepackage[switch]{lineno}
\usepackage{tabularx}
\usepackage{multirow}
\usepackage{latexsym}
\usepackage{amssymb}
\usepackage{bm}
\usepackage{tcolorbox}
\tcbuselibrary{breakable, skins}
\usepackage{listings}
\usepackage{dblfloatfix} 
\usepackage{placeins}

\newcommand{\valsd}[2]{#1$\pm$#2}

\newcommand{\bestvalsd}[2]{\textbf{#1$\pm$#2}}

\title{EvoCorps: An Evolutionary Multi-Agent Framework for Depolarizing Online Discourse}

\author{
Ning Lin$^{1\dagger}$ \and
Haolun Li$^{1\dagger}$ \and
Mingshu Liu$^{1\dagger}$ \and
Chengyun Ruan$^{1}$ \and
Kaibo Huang$^{1\ast}$ \and
Yukun Wei$^{1}$ \and
Zhongliang Yang$^{1\ast}$ \and
Linna Zhou$^{1}$
\affiliations
$^{1}$Beijing University of Posts and Telecommunications\\
\emails
\{linning, lhldudu, llms266, ruanchengyun815, huangkaibo, weiyukun, yangzl, zhoulinna\}@bupt.edu.cn
}

\begin{document}

\maketitle

\begingroup
\renewcommand{\thefootnote}{\fnsymbol{footnote}}
\footnotetext[2]{Equal contribution.}      
\footnotetext[1]{Corresponding authors.}   
\endgroup

\begin{abstract}
Polarization in online discourse erodes social trust and accelerates misinformation, yet technical responses remain largely diagnostic and post-hoc. Current governance approaches suffer from inherent latency and static policies, struggling to counter coordinated adversarial amplification that evolves in real-time. We present \textbf{EvoCorps}, an evolutionary multi-agent framework for proactive depolarization. EvoCorps frames discourse governance as a dynamic social game and coordinates roles for monitoring, planning, grounded generation, and multi-identity diffusion. A retrieval-augmented collective cognition core provides factual grounding and action--outcome memory, while closed-loop evolutionary learning adapts strategies as the environment and attackers change. We implement EvoCorps on the MOSAIC social-AI simulation platform for controlled evaluation in a multi-source news stream with adversarial injection and amplification. Across emotional polarization, viewpoint extremity, and argumentative rationality, EvoCorps improves discourse outcomes over an adversarial baseline, pointing to a practical path from detection and post-hoc mitigation to in-process, closed-loop intervention. The code is available at \url{https://github.com/ln2146/EvoCorps}.
\end{abstract}

\begin{figure}[!t]
   \centering
   \hspace*{-0.3cm}
   \includegraphics[width=1.1\linewidth]{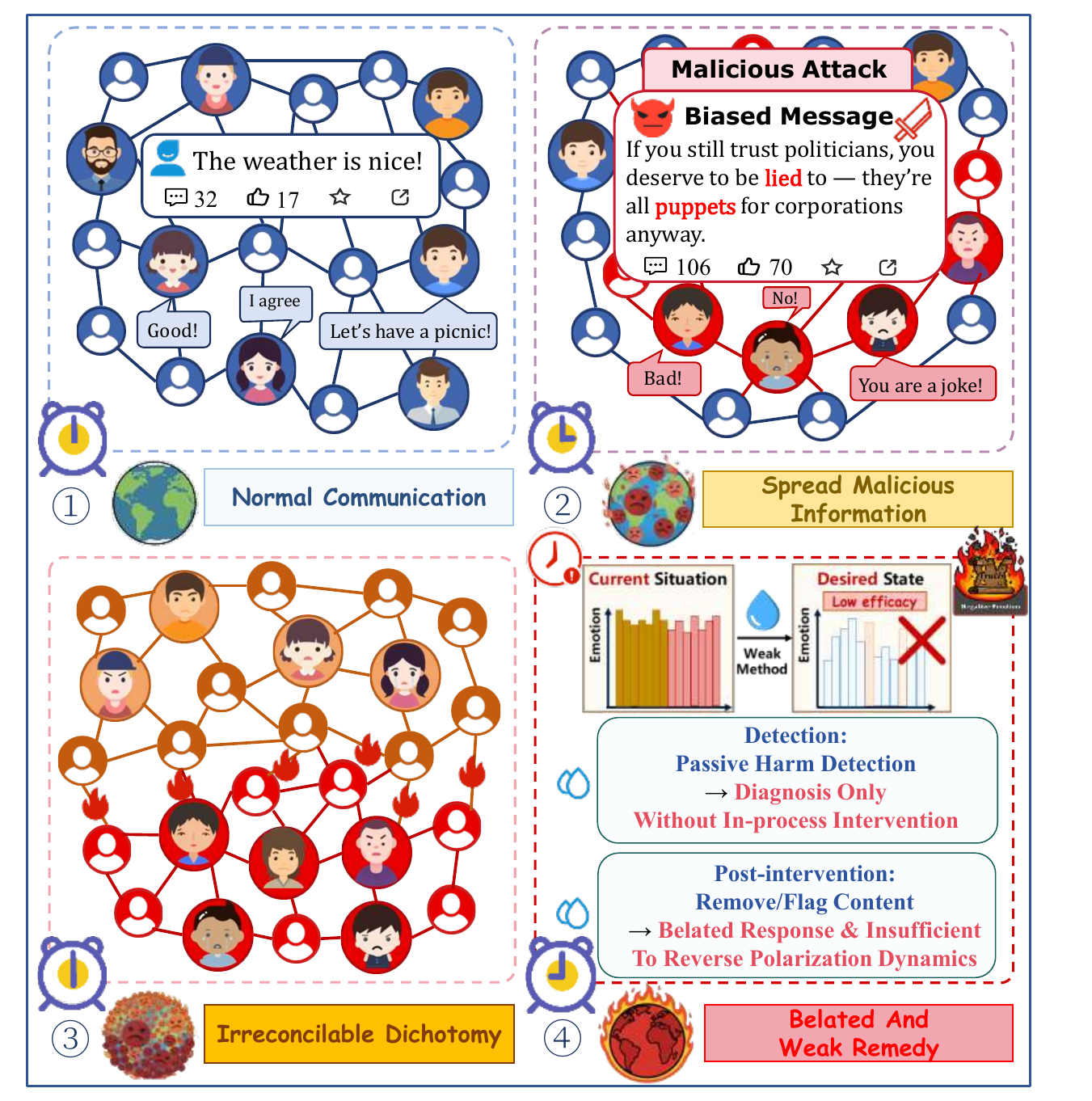}
   \caption{Online interaction evolves from normal user communication to malicious agents spreading harmful information, leading to amplified negative emotions and polarization. Existing harm detection and post-hoc intervention (e.g., content removal or labeling) occur only after diffusion and are insufficient to mitigate polarization.}
   \label{fig1}
\end{figure}

\section{Introduction}
Online social media platforms have become a central infrastructure for public communication, shaping how opinions, emotions, and narratives are formed at scale. Through algorithmic curation and homophilous network interactions, these platforms influence information exposure and social alignment. A consequence of these dynamics is the emergence of polarization in online discourse. Here, polarization refers to the segregation of users, content, and interactions into opposing camps with limited cross-cutting exposure; emotionally charged and one-sided narratives can become salient within homophilous communities, reinforcing group identities while marginalizing moderate viewpoints. \cite{loru2025ideology} This is a system-level outcome that can emerge from the interaction between homophily and engagement-optimized curation \cite{doi:10.1073/pnas.2023301118}. Prior work has shown that recommender systems can intensify echo chambers under strong homophily \cite{DBLP:journals/corr/abs-2112-00626}. Once entrenched, emotional reactions tend to dominate factual reasoning, rendering subsequent correction less effective.

Beyond these dynamics, contemporary online discourse is no longer shaped solely by user interactions. Accumulating evidence indicates that malicious and highly coordinated agents have been actively deployed in real-world platforms to influence public opinion and steer discourse trajectories. Recent studies demonstrate that such adversarial collectives can strategically exploit emotional contagion, algorithmic reinforcement, and coordinated multi-identity behaviors to reshape discourse dynamics at an early stage, often before existing governance mechanisms can respond \cite{feng-etal-2024-bot,ma-etal-2024-event}. Under these conditions, polarization is significantly accelerated through adversarial amplification, where coordinated agents inject and amplify emotionally provocative narratives across multiple identities, leveraging feedback loops in recommendation systems to rapidly diffuse content within homophilous communities and push discussions toward antagonistic camps (Figure~\ref{fig1}). In such settings, emotional signals frequently outpace factual corrections, allowing early frames to lock in and reducing the leverage of later moderation or fact-checking.

In many platforms, the dominant technical response to harmful or polarizing discourse remains diagnostic, where detection models and risk scoring can flag toxicity, stance extremity, or misinformation signals, but they do not by themselves execute timely, coordinated actions that change the trajectory of a live discussion. Building on detection, prior work has explored interventions such as fact-checking, content labeling, content removal or downranking, moderation workflows, and recommendation adjustments \cite{nakov-etal-2024-survey,vladika-matthes-2023-scientific,Habibi2024TheCM}. However, these mechanisms are often triggered after diffusion has already occurred or once polarization signals become salient, creating an inherent latency between early adversarial amplification and governance response.

Such latency reflects an asymmetric dynamic in governance, where defensive interventions are typically reactive, dispersed, and low-frequency, whereas adversarial collectives can act in a coordinated, anticipatory, and sustained manner. Under algorithmic feedback and homophilous interaction, emotionally charged narratives become path dependent; once amplified, engagement-driven exposure entrenches echo chambers \cite{FERRAZDEARRUDA2024111098}, and later corrections or isolated moderation actions have limited leverage. Moreover, episodic or static policies lack closed-loop adaptation to evolving adversarial behavior. These gaps motivate in-process governance mechanisms that operate throughout the interaction, coordinate complementary roles, and adapt.

To this end, we propose EvoCorps, an evolutionary multi-agent framework for proactive discourse depolarization under adversarial amplification. Inspired by the Stackelberg--Mean-Field Control (SMFC) paradigm \cite{HUANG20202237,Dayanikli2023AML}, EvoCorps models discourse governance as a dynamic social game in which an organized intervention team plans and executes strategies, while the broader user population responds at the population level. Operationally, EvoCorps instantiates this view as a closed-loop multi-agent system with an MMDP formulation and evolutionary learning that adapts strategies as discourse dynamics and adversarial tactics evolve.

Our main contributions are summarized as follows:

\begin{itemize}
    \item We formulate proactive depolarization as an in-process, closed-loop intervention problem under adversarial amplification, targeting the latency and non-adaptive nature of prevalent post-hoc interventions.
    \item We propose \textbf{EvoCorps}, combining role-specialized coordination with retrieval-augmented collective cognition and closed-loop evolutionary learning to adapt intervention strategies over time.
    \item We implement EvoCorps on MOSAIC and evaluate it under adversarial injection and coordinated amplification in a multi-source news stream. EvoCorps improves emotional polarization, viewpoint extremity, and argumentative rationality over adversarial and post-hoc baselines.
\end{itemize}
\section{Related Work}
\subsection{Agent-Based Malicious Information Diffusion}

Recent studies have revealed that malicious or coordinated agents can do more than spread falsehoods; by amplifying emotionally charged narratives, they can mobilize audiences into antagonistic camps, erode trust in institutions, and normalize hostile or conspiratorial framings.
Recent work shows that generative agents can propagate misinformation through coordinated liking and sharing \cite{liu-etal-2025-mosaic}, that AI--human collaboration can substantially extend the reach and persistence of sensational narratives \cite{ng2025appealscopemisinformationspread}, and that adaptive multi-agent coordination enables campaigns to evade moderation and sustain influence over time \cite{Ren2025WhenAG}.
Together, these findings suggest that adversarial collectives possess a systematic advantage in shaping discourse dynamics, as they exploit emotional contagion, algorithmic homophily, and coordinated amplification to drive fast, adaptive, and highly organized processes that often outpace reactive governance mechanisms.

\subsection{Research on Discourse Depolarization}

\begin{figure*}[!t]
    \includegraphics[width=1\textwidth]{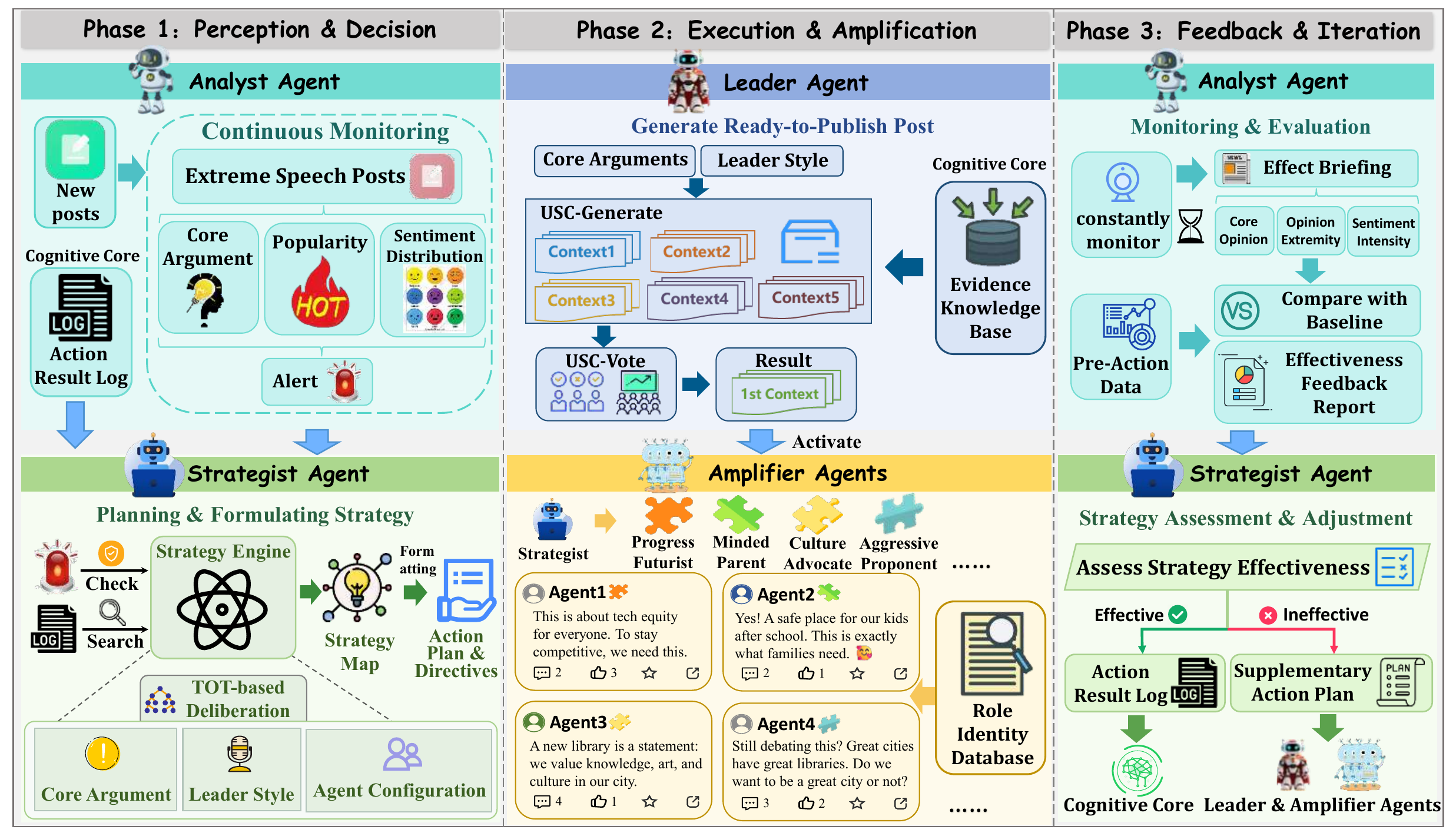}
    \caption{\textbf{Workflow of the closed-loop evolutionary system.} 
    In the perception and decision phase, an analyst agent monitors incoming posts and evaluates opinion stance, extremity, popularity, and sentiment to trigger warnings. A strategist agent then formulates core arguments, leader styles, and amplification scales. In the execution and amplification phase, the leader agent, supported by the USC mechanism, retrieves evidence, generates candidate contents, and selects optimal outputs via voting, while amplifier agents coordinate role-based identities to expand and propagate the content. In the feedback and iteration phase, an analyst agent evaluates intervention effects against baselines and produces structured reports, which are used by the strategist agent to guide strategy adjustment and support subsequent decision making.}
\label{fig2}
\end{figure*}

Most existing approaches to mitigating polarization in online discourse emphasize diagnostic capabilities, including measuring and detecting toxicity, misinformation, stance extremity, or biased narratives \cite{nakov-etal-2024-survey,vladika-matthes-2023-scientific,ma-etal-2024-event}. Recent studies explore polarization detection using self-supervised objectives and interaction features \cite{7f16220d6fe34a5d98cc318fd18e604a}, as well as large-scale analyses of polarized language and toxicity patterns in social media debates \cite{xu2025polarizedpatternslanguagetoxicity,ARORA2022121942}. While effective for identifying harmful content and auditing polarization indicators, such methods primarily provide diagnosis and do not directly specify how to steer the unfolding interaction process.

Building on these diagnostic signals, a range of interventions have been explored, most commonly in a post-hoc manner. Representative actions include account suspension or content takedown, content downranking, warning labels or interstitials, and attaching fact-checks or clarifications \cite{Habibi2024TheCM,nakov-etal-2024-survey}. These mechanisms can reduce further spread and exposure, yet they are typically triggered only after polarization signals become salient and are often executed as isolated actions.

As a result, these approaches remain largely reactive, failing to establish sustained feedback and strategy evolution over time, and thus fall short of effectively modeling and intervening in discourse dynamics as they unfold through interaction.

\section{Method}

\subsection{Problem Formulation and Environment Modeling}

We consider an online discourse environment consisting of a population of ordinary users interacting over an underlying network and consuming a time-indexed post stream $\{\mathcal{P}_t\}_{t=1}^T$. At each intervention round $t$, the environment reveals an observable discourse snapshot $o_t=(\mathcal{P}_t,\mathcal{C}_t)$ consisting of newly arrived posts and user comments. The intervention team then executes a joint intervention action $a_t$ following the closed-loop workflow in Figure~\ref{fig2}. The resulting interactions jointly determine the next-round discourse state under adversarial disturbances and stochastic shocks.

Rather than modeling all microscopic user states explicitly in the control layer, EvoCorps uses a compact mean-field summary of the environment. Specifically, we track average opinion extremity $v_t$ and aggregate sentiment $e_t$, estimated from the observation $o_t$ by the Analyst, as the control state for closed-loop decision-making.

To capture the closed-loop intervention process in a learnable form, we formalize one intervention round as one time step of a \textbf{Multi-Agent Markov Decision Process (MMDP)} 
\begin{equation}
\mathcal{M} = (\mathcal{N}, \mathcal{S}, \{\mathcal{A}^i\}_{i \in \mathcal{N}}, P, R),
\label{eq:MMDP}
\end{equation}
capturing the loop of perception, evaluation, decision, and execution across heterogeneous roles.  

\paragraph{Agent Set.}  
We define the set of agents as $\mathcal{N} = \{\text{Analyst}, \text{Strategist}, \text{Leader}, \text{Amplifier}\}$ with role-specific action spaces $\mathcal{A}^i$. Their functional roles are described in Sec.~3.3.  

\paragraph{State Space.}  
We use the mean-field state $s_t=(v_t,e_t)$, where $v_t$ denotes the average opinion extremity relative to a neutral baseline and $e_t$ denotes the aggregate sentiment level; $s_t$ is derived from the observation $o_t$ by the Analyst.  

\paragraph{Joint Action Space.}  
We distinguish between the full agent set $\mathcal{N}$ and the intervention subset $\mathcal{N}_c = \{\text{Strategist}, \text{Leader}, \text{Amplifier}\}$ that directly executes interventions; the Analyst performs observation and evaluation but is not included in the intervention action. At each time step, each agent $i \in \mathcal{N}_c$ selects an action $a_t^i \in \mathcal{A}^i$, and the joint intervention action is:  
\begin{equation}
a_t = (a_t^{\text{stg}}, a_t^{\text{ldr}}, a_t^{\text{amp}}) \in \mathcal{A}^{\text{stg}} \times \mathcal{A}^{\text{ldr}} \times \mathcal{A}^{\text{amp}} = \prod_{i \in \mathcal{N}_c} \mathcal{A}^i,
\label{eq:action_space}
\end{equation}
These components jointly define the intervention applied at time $t$.  

\paragraph{Reward Function.}  
The global cooperative reward is designed to reduce opinion extremity while increasing sentiment level:  
\begin{equation}
R(s_t, a_t) = -\lambda_1 \cdot \Delta v_t + \lambda_2 \cdot \Delta e_t,
\label{eq:reward}
\end{equation}
where $\Delta v_t = v_{t+1} - v_t$ and $\Delta e_t = e_{t+1} - e_t$.  
Here, $\lambda_1, \lambda_2 > 0$ balance opinion moderation against sentiment enhancement.  

\paragraph{System Dynamics.}  
State transitions are induced by the environment dynamics under the joint intervention:
\begin{equation}
s_{t+1} = f(s_t, a_t) + \epsilon_t,
\end{equation}
where $\epsilon_t$ captures exogenous shocks (e.g., misinformation bursts).  

\subsection{EvoCorps Framework Overview}

Figure~\ref{fig2} summarizes a closed-loop workflow with three phases, supported by a retrieval-augmented cognition core and role-specialized agents.  
\textbf{Perception and Decision} produces a structured plan from live discourse signals and historical memory, primarily driven by the Analyst and Strategist.  
\textbf{Execution and Amplification} generates grounded responses and diffuses them through role-based identities, executed by the Leader and Amplifier.  
\textbf{Feedback and Iteration} evaluates intervention effects and updates memory; the Analyst produces structured reports and the Strategist uses them to revise subsequent rounds.  
These phases are system-level modules, while role-specialized agents implement the modules and produce the concrete intervention outputs.

\subsection{Dynamic Game Team with Role Coordination}

Unlike prior frameworks that treat agents as homogeneous entities, EvoCorps models interventions as the coordinated strategies of heterogeneous teams. This design aligns with real-world online governance, where discourse is shaped not by isolated individuals but by organized groups such as fact-checking alliances, advocacy organizations, or community networks, each dividing responsibilities across specialized roles.

EvoCorps defines four functional roles with complementary responsibilities: 

\textbf{Analyst.} Continuously monitors discourse and extracts core claims, viewpoint extremity, current popularity, and sentiment polarity. When polarization risks are detected, the analyst issues structured alerts to trigger coordinated intervention. After interventions, the analyst also evaluates outcomes against baseline states, providing feedback for subsequent iterations. 

\textbf{Strategist.} Formulates intervention strategies by combining current alerts with historical experience stored in the Action--Outcome Memory. The strategist determines not only which narratives to counter or reinforce, but also the leader's rhetorical style and the composition of amplifier identities, issuing structured directives to guide coordinated action. 

\textbf{Leader.} Generates high-quality, persuasive content aligned with the strategist's plan. To ensure factual grounding and stylistic diversity, the leader retrieves arguments from the Evidence Knowledge Base, produces multiple candidate drafts, and applies reflection-and-voting mechanisms to select the most effective output. 

\textbf{Amplifier.} Disseminates leader's outputs across the simulated environment by adopting diverse role identities (e.g., ordinary users, topical experts, community figures). Operating in parallel, amplifiers generate varied yet coherent responses, rapidly amplifying evidence-grounded narratives and signaling broader, evidence-based alignment in the discussion.

Within a round, role-specific decisions are made sequentially (Analyst $\rightarrow$ Strategist $\rightarrow$ Leader $\rightarrow$ Amplifier), while their outputs jointly constitute the intervention action and determine the next-state transition (Figure~\ref{fig2}).

This role-specialized pipeline, visualized in Figure~\ref{fig2}, enables collective intervention via division of labor beyond prior homogeneous-agent settings.

\subsection{Retrieval-Augmented Collective Cognition Core and Evolutionary Learning}

LLM-driven multi-agent systems often suffer from inconsistency and memory loss, which hinders long-horizon coordination. EvoCorps addresses this with a \textit{retrieval-augmented collective cognition core} that couples external knowledge with action--outcome memory. The intervention process is formalized as a closed-loop system: at time step $t$, the joint action is sampled from a state-feedback policy conditioned on the environment state and the cognition core,
\begin{equation}
a_t \sim \pi(a_t \mid s_t, K_t, M_t),
\label{eq:policy}
\end{equation}
where $s_t = (v_t, e_t)$ is the mean-field state of the discourse field, $K_t$ the evidence knowledge base, and $M_t$ the action--outcome memory. The policy is instantiated through the role-coordinated team (see Sec.~3.3), producing the joint intervention action defined in Eq.~\ref{eq:action_space}.

The learning objective maximizes cumulative reward,
\begin{equation}
J = \mathbb{E}_{\pi} \left[\sum_{t=1}^T R(s_t, a_t)\right],
\label{eq:objective}
\end{equation}
where $R(s_t, a_t)$ is defined in Eq.~\ref{eq:reward}. Unlike gradient-based RL, optimization is achieved by selectively evolving $K_t$ and $M_t$ via reward feedback rather than updating parameters $\theta$.

\paragraph{Evidence Knowledge Base.}
The repository stores arguments with persuasiveness scores, continuously retrieved from trusted sources. At time step $t$,
\begin{equation}
K_t = \{(k_i, p_i)\}_{i=1}^n, \quad p_i \in [0,1],
\end{equation}
where $k_i$ is a fact or argument and $p_i$ its persuasiveness score. A retrieved item $(k', p')$ is incorporated if
\begin{equation}
\text{relevance}(k', K_t) > \delta, \quad \delta \in (0,1),
\end{equation}
yielding $K_{t+1} = K_t \cup \{(k', p')\}$.

After each round, successful arguments are reinforced while ineffective ones are down-weighted:
\begin{equation}
p_i \leftarrow p_i + \eta \cdot R(s_t, a_t) \cdot \mathbb{I}[k_i \in \mathcal{K}(a_t)],
\label{eq:score_update}
\end{equation}
where $\eta > 0$ is the learning rate and $\mathbb{I}[k_i \in \mathcal{K}(a_t)]$ indicates whether $k_i$ is selected in action $a_t$, with $\mathcal{K}(a_t)$ the selected knowledge set.

\paragraph{Action--Outcome Memory.}
The cognition core records each action and its observed effect:
\begin{equation}
M_t = \{(a_i, o_i, r_i)\}_{i=1}^{t-1},
\end{equation}
where $a_i$ is an action, $o_i$ the resulting observation snapshot, and $r_i = R(s_i, a_i)$ the associated reward with $s_i$ the mean-field state at step $i$. The log supports strategic evaluation.

To maintain adaptability, a new tuple is incorporated only when the reward exceeds a threshold $\epsilon$:
\begin{equation}
R(s_t, a_t) > \epsilon, \quad \epsilon \in (0,1),
\end{equation}
and the updated memory becomes
\begin{equation}
M_{t+1} = M_t \cup \{(a_t, o_t, r_t)\}.
\end{equation}

\paragraph{Evolutionary Learning Dynamics.}
Reward signals and memory updates form a closed-loop evolutionary learning system: policy improvement is realized through the score updates in Eq.~\ref{eq:score_update} and selective retention, with $\pi(a \mid s, K_t, M_t)$ conditioned on memory rather than a parameterized $\pi_\theta(a \mid s)$.

This design extends depolarization into a reinforcement-style multi-agent evolutionary process: successful knowledge and action patterns are retained while ineffective ones are forgotten, enabling adaptive long-horizon coordination without parameter updates.

\section{Experiments}
\subsection{Experimental Setup}

\begin{table*}[!t]
\centering
\small
\renewcommand{\arraystretch}{1.05}

\begin{tabular*}{\textwidth}{@{\extracolsep{\fill}}clcccccc}
\toprule
\multirow{2}{*}{$t$} & \multirow{2}{*}{Case} &
 \multicolumn{2}{c}{Emotion (\%)} &
 \multicolumn{1}{c}{Viewpoint (\%)} &
 \multicolumn{3}{c}{Argument (\%)} \\
\cmidrule(lr){3-4}\cmidrule(lr){5-5}\cmidrule(lr){6-8}
 & 
 & Sentiment $\uparrow$
 & Toxicity $\downarrow$
 & Extremity $\downarrow$
 & AQS $\uparrow$
 & Fallacy $\downarrow$
 & Evidence $\uparrow$ \\
\midrule

\multirow{4}{*}{1}
 & Case~1 & \bestvalsd{57.0}{4.1} & \bestvalsd{3.3}{0.7} & \valsd{19.9}{4.2} & \bestvalsd{38.2}{2.6} & \valsd{20.3}{8.1} & \valsd{24.6}{1.8} \\
 & Case~2 & \valsd{56.3}{0.6} & \valsd{3.7}{0.5} & \valsd{19.5}{1.8} & \valsd{38.1}{1.7} & \valsd{16.0}{3.5} & \bestvalsd{25.5}{1.1} \\
 & Case~3 & \valsd{55.0}{5.3} & \valsd{3.6}{0.8} & \valsd{20.6}{2.3} & \valsd{37.1}{0.4} & \valsd{15.1}{3.7} & \valsd{25.4}{0.3} \\
 & Case~4 & \valsd{56.8}{3.3} & \valsd{3.6}{0.3} & \bestvalsd{18.9}{0.6} & \valsd{36.2}{0.1} & \bestvalsd{10.2}{4.9} & \valsd{24.3}{0.5} \\
\midrule
\midrule

\multirow{4}{*}{5}
 & Case~1 & \valsd{41.2}{2.5} & \bestvalsd{4.7}{0.3} & \valsd{34.6}{3.2} & \valsd{38.7}{0.3} & \valsd{35.1}{7.7} & \valsd{25.2}{1.1} \\
 & Case~2 & \valsd{42.1}{3.5} & \valsd{6.1}{0.7} & \valsd{31.1}{2.9} & \valsd{41.2}{0.3} & \valsd{20.1}{4.6} & \valsd{26.8}{0.6} \\
 & Case~3 & \valsd{45.7}{2.5} & \valsd{5.1}{0.7} & \valsd{28.7}{2.2} & \bestvalsd{41.4}{0.8} & \valsd{16.6}{3.4} & \valsd{27.1}{0.7} \\
 & Case~4 & \bestvalsd{49.1}{3.3} & \valsd{5.2}{0.5} & \bestvalsd{24.6}{1.3} & \valsd{40.9}{1.5} & \bestvalsd{14.3}{3.9} & \bestvalsd{27.9}{1.2} \\
\midrule
\midrule

\multirow{4}{*}{10}
 & Case~1 & \valsd{36.2}{4.9} & \valsd{8.3}{0.3} & \valsd{40.2}{2.7} & \valsd{38.9}{0.0} & \valsd{40.1}{2.4} & \valsd{25.4}{0.8} \\
 & Case~2 & \valsd{33.1}{3.7} & \valsd{8.0}{1.2} & \valsd{37.8}{3.1} & \valsd{42.1}{0.6} & \valsd{28.7}{5.4} & \valsd{27.2}{0.9} \\
 & Case~3 & \valsd{37.4}{4.2} & \bestvalsd{6.7}{1.0} & \valsd{34.5}{3.1} & \valsd{42.2}{1.3} & \valsd{24.1}{5.8} & \valsd{27.7}{0.9} \\
 & Case~4 & \bestvalsd{42.5}{3.1} & \valsd{7.0}{0.7} & \bestvalsd{29.5}{2.5} & \bestvalsd{43.8}{0.8} & \bestvalsd{20.6}{4.4} & \bestvalsd{30.1}{0.5} \\
\midrule
\midrule

\multirow{4}{*}{20}
 & Case~1 & \valsd{36.2}{4.6} & \valsd{7.8}{0.1} & \valsd{41.8}{2.6} & \valsd{41.9}{0.3} & \valsd{37.1}{2.1} & \valsd{27.2}{0.9} \\
 & Case~2 & \valsd{27.1}{1.7} & \valsd{9.2}{0.6} & \valsd{43.1}{2.2} & \valsd{42.3}{0.1} & \valsd{36.0}{2.0} & \valsd{27.0}{0.4} \\
 & Case~3 & \valsd{31.1}{3.1} & \valsd{8.3}{1.0} & \valsd{40.0}{2.3} & \valsd{42.4}{1.2} & \valsd{30.9}{5.0} & \valsd{27.7}{1.0} \\
 & Case~4 & \bestvalsd{39.9}{1.8} & \bestvalsd{7.6}{0.7} & \bestvalsd{31.2}{1.8} & \bestvalsd{44.9}{0.8} & \bestvalsd{22.6}{3.0} & \bestvalsd{28.2}{0.9} \\
\midrule
\midrule

\multirow{4}{*}{30}
 & Case~1 & \valsd{33.4}{3.0} & \valsd{8.2}{0.1} & \valsd{44.7}{0.8} & \valsd{43.0}{0.1} & \valsd{36.1}{2.2} & \valsd{27.9}{0.6} \\
 & Case~2 & \valsd{25.1}{1.3} & \valsd{9.7}{0.6} & \valsd{45.1}{1.7} & \valsd{42.4}{0.2} & \valsd{37.2}{2.0} & \valsd{27.0}{0.3} \\
 & Case~3 & \valsd{29.0}{3.1} & \valsd{8.7}{1.0} & \valsd{41.8}{2.1} & \valsd{43.1}{1.4} & \valsd{32.2}{4.4} & \valsd{29.0}{1.4} \\
 & Case~4 & \bestvalsd{39.2}{2.0} & \bestvalsd{7.9}{0.5} & \bestvalsd{31.1}{2.1} & \bestvalsd{45.4}{0.6} & \bestvalsd{21.7}{2.8} & \bestvalsd{31.3}{0.5} \\
\bottomrule
\end{tabular*}

\caption{\textbf{User-level effectiveness at selected time steps for four experimental conditions.}
Case~1: benign baseline with no adversary and no intervention.
Case~2: adversarial amplification with no protection.
Case~3: post-hoc intervention baseline.
Case~4: EvoCorps with proactive coordination.
Metrics evaluate emotional polarization, viewpoint extremity, and argumentative rationality.}
\label{tab:effectiveness_users}
\end{table*}

\begin{figure*}[!t]
    \centering
    \hspace*{-0.8cm}
    \includegraphics[width=1.05\textwidth]{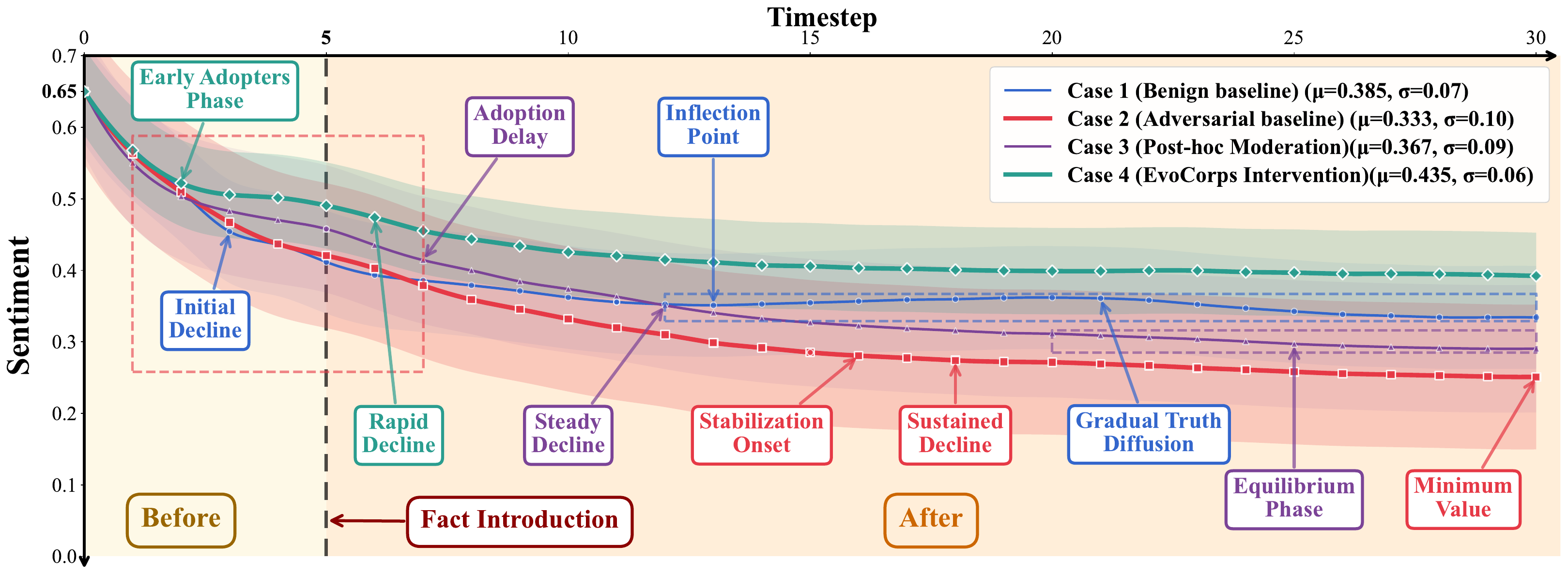}
\caption{\textbf{Sentiment trajectories over time under Case~1/2/3/4.} The dashed line marks clarification at $t{=}5$. Case~2 continues to decline, Case~3 partially mitigates, and Case~4 declines more slowly and stabilizes relative to Case~2/3.}
\label{fig3}
\end{figure*}

We run our experiments on top of the MOSAIC social-AI simulation platform \cite{liu-etal-2025-mosaic}, extending it with a multi-source news stream, adversarial injection and amplification, and a coordinated intervention team.

\paragraph{News stream and adversarial setting.}
We build a multi-source news stream from NELA-GT-2021 and the COVID-19 Fake News Dataset \cite{gruppi2023nelagt2022largemultilabellednews,Patwa_2021}. For a subset of real news items, we use an LLM to generate extremized variants as adversarial stimuli and inject the corresponding factual clarifications after a predetermined delay (4 time steps in our experiments). Malicious agents preferentially amplify adversarial sources to sustain polarization pressure.

\paragraph{Argumentation knowledge base.}
We build an Evidence Knowledge Base by integrating IBM Debater arguments \cite{mirkin-etal-2018-recorded} with relevant materials retrieved via the Wikipedia API.

\paragraph{Agents and evaluation configuration.}
All users and agents are instantiated from LLM-enriched Nemotron-Personas \cite{nvidia/Nemotron-Personas-USA}. We simulate 50 ordinary users for $T{=}30$ time steps under a fixed news stream; to increase heterogeneity, ordinary users sample between \texttt{gpt-4.1-nano} and \texttt{gemini-2.0-flash} \cite{openai_gpt41nano_2025,google_gemini20flash_2025} (Appendix Fig.~\ref{fig5}). For adversarial settings, we introduce malicious agents that preferentially target and amplify adversarial sources; we keep the news stream, population, horizon, and adversary fixed across Case~2/3/4 and vary only the defense mechanism.



\paragraph{Experimental conditions.}
Based on this setup, we define four conditions (Case~1/2/3/4) for the main comparison:
\begin{itemize}
    \item \textbf{Case~1 (Benign Baseline).} No adversary and no intervention, with discourse evolving organically from ordinary users.
    \item \textbf{Case~2 (Adversarial Propagation).} Malicious agents amplify adversarial sources, with no protection.
    \item \textbf{Case~3 (Post-hoc Moderation).} On top of Case~2, an LLM-based reviewer triggers fact-checking and takedown \cite{nakov-etal-2024-survey,vladika-matthes-2023-scientific,Habibi2024TheCM,Li2024LLMsasJudgesAC}.
    \item \textbf{Case~4 (EvoCorps Intervention).} On top of Case~2, EvoCorps performs proactive, role-coordinated intervention.
\end{itemize}

For each condition, we select posts with more than 50 comments and compute all metrics on the \emph{ordinary-user subset} to assess user-level outcomes.

\subsection{Evaluation Metrics}

\begin{table*}[t]
\centering
\small
\setlength{\tabcolsep}{2.8pt}
\renewcommand{\arraystretch}{1.08}

\begin{tabular*}{\textwidth}{@{\extracolsep{\fill}}lcccccc}
\toprule
\multirow{2}{*}{\textbf{Model}} &
 \multicolumn{2}{c}{\textbf{Emotion (\%)}} &
 \multicolumn{1}{c}{\textbf{Viewpoint (\%)}} &
 \multicolumn{3}{c}{\textbf{Argument (\%)}} \\
\cmidrule(lr){2-3}\cmidrule(lr){4-4}\cmidrule(lr){5-7}
 & \textbf{Sentiment $\uparrow$}
 & \textbf{Toxicity $\downarrow$}
 & \textbf{Extremity $\downarrow$}
 & \textbf{AQS $\uparrow$}
 & \textbf{Fallacy $\downarrow$}
 & \textbf{Evidence $\uparrow$} \\
\midrule
Full Model      & \bestvalsd{39.2}{2.0} & \bestvalsd{7.9}{0.5} & \bestvalsd{31.1}{2.1} & \bestvalsd{45.4}{0.6} & \bestvalsd{21.7}{2.8} & \bestvalsd{31.3}{0.5} \\
w/o Analyst     & \valsd{33.9}{5.3}     & \valsd{8.5}{0.6}     & \valsd{33.5}{2.9}     & \valsd{44.5}{1.7} & \valsd{26.8}{4.8}     & \valsd{30.7}{0.9} \\
w/o Strategist  & \valsd{30.9}{1.2}     & \valsd{8.6}{0.3}     & \valsd{41.0}{1.4}     & \valsd{43.7}{0.8}     & \valsd{32.9}{2.2}     & \valsd{28.6}{0.4} \\
w/o Leader      & \valsd{28.4}{2.0}     & \valsd{8.4}{0.4}     & \valsd{41.9}{2.2}     & \valsd{42.4}{0.8}     & \valsd{34.9}{4.8}     & \valsd{27.7}{0.3} \\
w/o Amplifiers  & \valsd{26.9}{4.9}     & \valsd{9.5}{1.2}     & \valsd{43.2}{4.6}     & \valsd{41.8}{2.1}     & \valsd{39.1}{7.7}     & \valsd{27.1}{1.7} \\
\bottomrule
\end{tabular*}

\caption{\textbf{EvoCorps role ablation study at $t{=}30$.}}
\label{tab:ablation}
\end{table*}

To assess EvoCorps' depolarization capability, we evaluate emotional polarization, viewpoint extremity, and argumentative rationality; all conditions use identical graders and settings, so we report relative differences following common practice in LLM-based evaluations \cite{lin-chen-2023-llm,Byun2025LLMasaGraderPI,Li2024LLMsasJudgesAC}. Evaluation protocol and prompt templates are provided in Appendix~\ref{app:metric_details}.

\paragraph{Emotional Polarization.}
We score ordinary-user utterances on a five-level sentiment scale (linearly mapped to $[0,1]$) and compute toxicity via the Perspective API \cite{luong-etal-2024-realistic,plaza-del-arco-etal-2024-emotion}. Higher sentiment and lower toxicity indicate better emotional regulation.

\paragraph{Viewpoint Extremity.}
We label viewpoint extremity into five ordered levels (linearly mapped to $[0,1]$) with deterministic decoding and report the mean over ordinary-user comments \cite{nakov-etal-2024-survey,doi:10.1177/14614448221117484}. Lower values indicate more moderate viewpoints.

\paragraph{Argumentative Rationality.}
We grade stance-bearing comments using fixed rubrics to obtain argument quality (AQS) \cite{wachsmuth-etal-2024-argument}, fallacy rate \cite{yeh-etal-2024-cocolofa}, and evidence usage \cite{liu-etal-2024-empirical}. Higher AQS/evidence and lower fallacy indicate stronger argumentative rationality.

\subsection{Evaluation Results}

We evaluate user-level outcomes under identical simulation and adversarial configurations, varying only the intervention mechanism across the four cases defined in the experimental setup. Table~\ref{tab:effectiveness_users} reports snapshot metrics at $t \in \{1,5,10,20,30\}$, and Figure~\ref{fig3} plots sentiment trajectories over time; the first clarification is introduced at $t{=}5$.

\paragraph{Trajectory change under adversarial amplification.}
Under adversarial amplification, Case~2 fails to recover after clarification, with sentiment continuing to deteriorate while extremity rises, suggesting post-hoc corrections arrive too late after amplification takes hold. Table~\ref{tab:effectiveness_users} and Figure~\ref{fig3} show EvoCorps as the only setting that changes this trajectory. The divergence appears soon after $t{=}5$, with sentiment stabilizing relative to Case~2/3 and extremity staying lower through $t{=}30$.

\paragraph{Baseline contrast and endpoints.}
Using Case~2 as the primary adversarial baseline, Case~3 provides limited improvements. Toxicity decreases to $8.7$ by $t{=}30$, yet viewpoint extremity still rises to $41.8$, indicating post-hoc moderation can dampen tone while leaving entrenched viewpoints largely unchanged. Case~1 serves as a no-adversary reference rather than an upper bound, since polarization can emerge organically without intervention, with extremity reaching $44.7$ at $t{=}30$. By contrast, Case~4 is the only setting that suppresses extremity to $31.1$ while improving evidence-grounded rationality. It achieves $31.3$ evidence usage and $21.7$ fallacy, with toxicity comparable overall to Case~3 but lower at later steps.

\paragraph{Metric emphasis and complementarity.}
Differences between Case~3 and Case~4 are most pronounced in sentiment, extremity, and argument quality rather than toxicity. EvoCorps moderates viewpoints early, with extremity at $24.6$ versus $28.7$ at $t{=}5$. The gap widens over time, reaching $31.2$ versus $40.0$ at $t{=}20$ and $31.1$ versus $41.8$ at $t{=}30$. Sentiment follows a similar trend, with gaps of $42.5$ versus $37.4$ at $t{=}10$ and $39.2$ versus $29.0$ at $t{=}30$. Evidence usage and fallacy rates further favor Case~4 at later stages, as shown in Table~\ref{tab:effectiveness_users}. Toxicity differences are smaller and not consistently in Case~4's favor at early steps. Together, these metrics capture complementary aspects of discourse dynamics: sentiment and extremity reflect affective and ideological shifts, while argument quality reflects discourse rationality.

\begin{figure}[t]
    \centering
    \includegraphics[width=1.0\columnwidth]{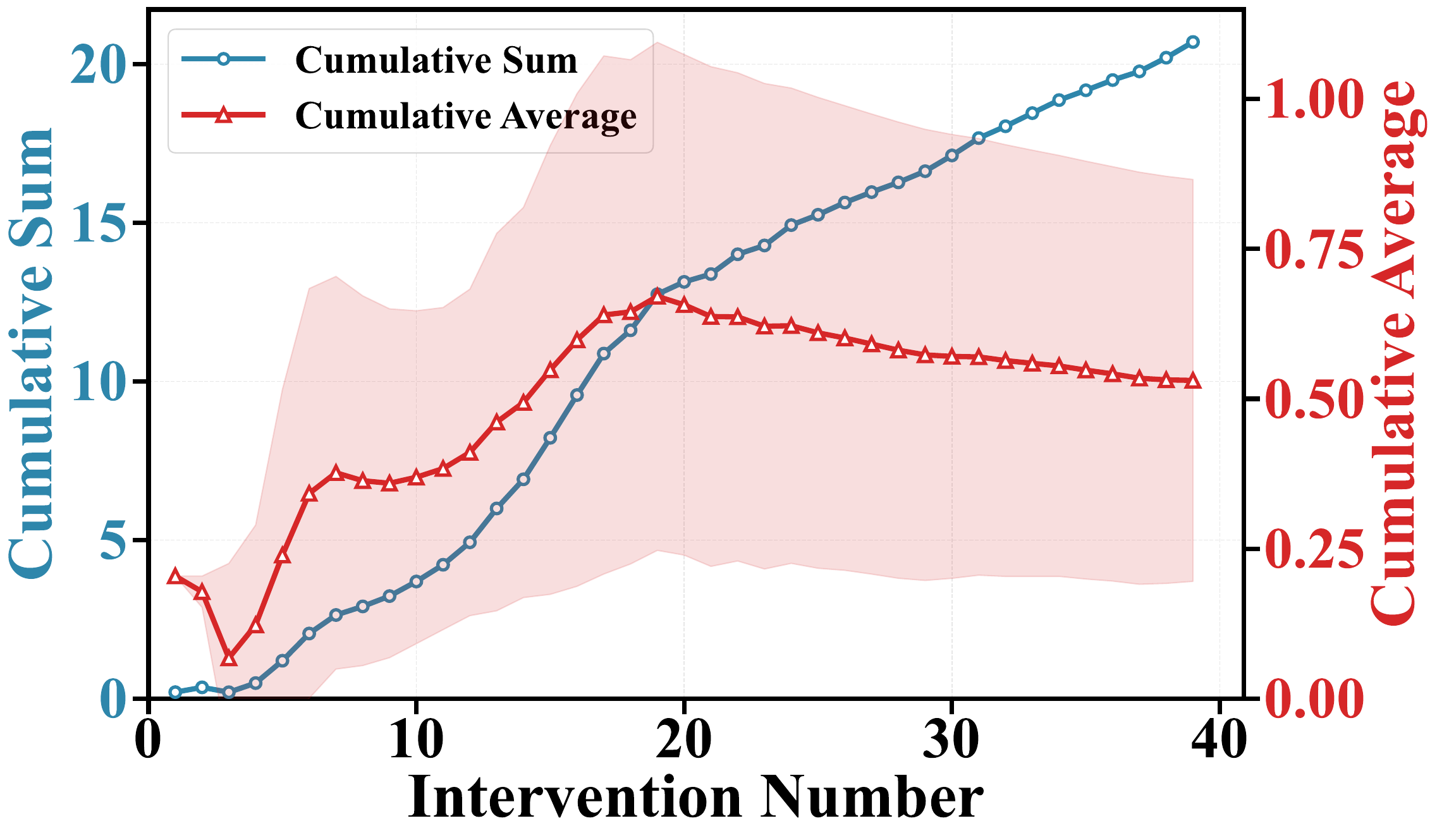}
\caption{\textbf{Reward trajectories across intervention rounds.} Reward is defined in Eq.~\ref{eq:reward} from changes in sentiment and viewpoint extremity. The cumulative sum grows steadily, while the cumulative average rises in early rounds and then stabilizes, reflecting diminishing marginal gains as the discourse state becomes more stable.}
\label{fig7}
\end{figure}

\paragraph{Reward dynamics.}
To connect our closed-loop objective to observed behavior, Figure~\ref{fig7} plots reward trajectories computed from Eq.~\ref{eq:reward}. The cumulative reward increases across intervention rounds, while the cumulative average rises early and then stabilizes, consistent with larger gains when the discourse is most volatile and smaller marginal improvements once the state becomes more stable.

\subsection{Ablation Study}

\paragraph{Counter-amplification is the dominant lever.}
Table~\ref{tab:ablation} reports single-role ablations at $t{=}30$ under the same adversarial setting as Case~4. Removing \textbf{Amplifiers} causes the largest regressions across all dimensions. Viewpoint extremity increases by $12.1$ points from $31.1$ to $43.2$. The fallacy rate nearly doubles from $21.7$ to $39.1$, and evidence usage drops from $31.3$ to $27.1$. This identifies counter-amplification as the primary bottleneck, since grounded responses without sufficient placement and reach cannot shift ordinary-user trajectories under coordinated attacks.

\paragraph{Planning and leadership translate diffusion into rational gains.}
Removing the \textbf{Strategist} or \textbf{Leader} sharply weakens viewpoint moderation, with extremity increasing from $31.1$ to $41.0/41.9$ and fallacies rising from $21.7$ to $32.9/34.9$. The w/o Strategist variant shows a similar collapse, with sentiment dropping to $30.9$ (vs.\ $39.2$), extremity rising to $41.0$, and fallacy to $32.9$. These results indicate that high-quality generation alone is insufficient; strategic coordination and execution are required to convert diffusion into evidence-grounded influence, and without them the system degenerates into situation-agnostic posting akin to fixed-policy bots.

\paragraph{Analyst stabilizes timing and reduces volatility.}
Removing the \textbf{Analyst} yields moderate degradations across metrics and a clear rise in variance. Sentiment drops from $39.2$ to $33.9$, while toxicity rises from $7.9$ to $8.5$. Extremity increases from $31.1$ to $33.5$, and argumentative quality softens, with AQS $45.4 \rightarrow 44.5$, fallacy $21.7 \rightarrow 26.8$, and evidence $31.3 \rightarrow 30.7$. The largest shift is in volatility, with sentiment standard deviation increasing from $2.0$ to $5.3$. This suggests the analyst stabilizes intervention timing, reducing oscillatory over- or under-reaction and yielding steadier trajectories even when mean shifts are modest.

\section{Conclusion}

This paper proposes \textbf{EvoCorps}, an evolutionary multi-agent framework that shifts online discourse governance from detection and post-hoc mitigation to in-process, closed-loop intervention under adversarial amplification. Built on MOSAIC with a multi-source news stream, EvoCorps reshapes the post-shock trajectory and improves emotional polarization, viewpoint extremity, and argumentative rationality compared with both an adversarial baseline and a post-hoc baseline based on LLM review and takedown. Role ablations identify counter-amplification as a key bottleneck and show that grounded planning and generation are important contributors to robust gains.

\newpage
\bibliographystyle{named}
\bibliography{ijcai26}

\clearpage

\appendix

\section{Limitations}
Despite its strong performance in controlled simulations, EvoCorps has several limitations. Our experiments are conducted in a text-based simulation environment built on curated datasets such as NELA-GT, IBM Debater, and Nemotron-Personas. While this enables controlled analysis, it does not capture multimodal signals like images or memes that often influence real-world polarization. As a result, the external validity of our findings may be limited when applied to more complex online settings.

In real-world deployment, intervention systems face more complex human–machine interaction dynamics than those modeled here. In particular, users may exhibit psychological reactance, resisting interventions simply because they originate from an automated system. This factor is not explicitly modeled in our current framework and motivates future human-in-the-loop evaluations to assess how user perception and trust affect intervention effectiveness.

EvoCorps relies on large language models for agent behavior and partial evaluation, which introduces known issues such as hallucinations and latent biases. Large language models may also exhibit tendencies towards compliance. However, since all experimental conditions are evaluated using identical prompts, models, and grading procedures, such biases are controlled across groups. Accordingly, our results should be interpreted primarily in terms of relative performance and the robustness of qualitative conclusions rather than absolute guarantees.

Practical deployment involves additional constraints, including platform policies, transparency requirements, and deployment costs, which are not explicitly modeled in this study and require further validation in more realistic environments.

\section{Ethics Statement}
This work investigates mechanisms for online discourse depolarization in a simulated environment, utilizing publicly available datasets and synthetic agent interactions. It does not involve experiments with human subjects and does not collect or process personally identifying information. The primary goal of this research is to advance understanding of coordinated intervention mechanisms for platform governance, rather than to develop or deploy deceptive influence campaigns.

EvoCorps is framed as a governance-assistance approach for online platforms facing coordinated and malicious activities such as disinformation campaigns or adversarial manipulation. In such settings, platform governance actors may themselves require coordinated capabilities and stylistic diversity to respond effectively and proportionately. Our study therefore examines coordination and response diversity as governance mechanisms, not as tools for artificial consensus formation or manipulation.

We explicitly oppose the use of deceptive strategies in any real-world deployment. Although our simulations introduce diverse agent personas to explore theoretical boundaries of influence dynamics, any practical application must adhere strictly to principles of transparency and accountability. Automated agents should be clearly identified as AI-based assistants or governance tools, such as certified fact-checking bots, and must not impersonate human users or conceal their artificial nature.

Any deployment of systems inspired by this work should be integrated with existing platform governance processes and subject to platform-specific policies, transparency requirements, and continuous auditing. Such safeguards are necessary to mitigate unintended harms, including disparate impacts, erosion of user trust, or errors arising from automated judgments. The intended use of EvoCorps is to support responsible, transparent, and accountable governance interventions, rather than to mislead users or manufacture false consensus.

\section{Evaluation Metric Details}
\label{app:metric_details}

\subsection{Evaluation Protocol and Aggregation}

Unless otherwise stated, we compute all content-based metrics on the \emph{ordinary-user subset} (records with agent\_type = \texttt{normal}) from data/scenario\_*/post-*/comments.jsonl. Following the main setup, we restrict evaluation to posts with more than 50 comments and report snapshot values at selected time steps $t\in\{1,5,10,20,30\}$. For each snapshot $t$, we aggregate over all eligible ordinary-user comments with \texttt{time\_step}$\le t$, where \texttt{time\_step} is the 1-indexed simulation step stored in the dataset.

LLM-based metrics use fixed prompts and deterministic decoding (e.g., temperature $=0$ where applicable), and we keep the grader model/API configuration identical across experimental conditions. We skip empty texts; invalid grader outputs are treated as missing for aggregation.

\subsection{Emotional Polarization}

\paragraph{Sentiment.} We label each ordinary-user comment with \texttt{gpt-4o-mini} into five classes \{\emph{Very Negative, Negative, Neutral, Positive, Very Positive}\} and map them to scores $\{0,0.25,0.5,0.75,1\}$. We report sentiment as $100\cdot\frac{1}{|C|}\sum_{c\in C} s_{\text{sent}}(c)$.
\begin{tcolorbox}[breakable, title=Sentiment Grader Prompt]
\small\raggedright
System: You are a helpful sentiment classifier.\\
User:\\
You are a sentiment analysis model.\\
Classify the following text into exactly one of the five categories: ["Very Negative", "Negative", "Neutral", "Positive", "Very Positive"]\\
Text: \textless TEXT\textgreater\\
Return only the label.\\
\end{tcolorbox}

\paragraph{Toxicity.} We score toxicity using the Perspective API, obtaining continuous toxicity probabilities in $[0,1]$ per comment. We report toxicity as $100\cdot\frac{1}{|C|}\sum_{c\in C} s_{\text{tox}}(c)$ (lower is better), with optional auxiliary statistics such as the percentage above a high-toxicity threshold.

\subsection{Viewpoint Extremity}

\paragraph{Viewpoint Extremity.}
We quantify viewpoint extremity by labeling each post and comment into one of five classes: \{\textit{Very Moderate}, \textit{Moderate}, \textit{Neutral}, \textit{Extreme}, \textit{Very Extreme}\} using \texttt{gpt-4o-mini} with deterministic decoding, \texttt{temperature}$=0$. We map labels to scalar scores in $[0,1]$ via $\{0,0.25,0.5,0.75,1\}$, respectively.
For a snapshot with eligible ordinary-user comments $C$, we report the mean extremity, scaled to $[0,100]$:
\[
\begin{split}
\text{ViewpointExtremity}
= 100 \cdot \frac{1}{|C|}\sum_{c\in C} s_{\text{ve}}(c), \\
s_{\text{ve}}(c)\in\{0,0.25,0.5,0.75,1\}.
\end{split}
\]
To characterize dispersion, we compute the score variance $\mathrm{Var}_{c\in C}[s_{\text{ve}}(c)]$. We additionally compute the label distribution $p_k$ over the five classes and its Shannon entropy $H=-\sum_k p_k \log p_k$ to capture whether viewpoints are concentrated at extreme levels or spread across moderate and neutral levels.
As an auxiliary consistency check, we compare the LLM label with the precomputed \texttt{predicted\_label} stored in the data. Treating the five labels as ordered levels, we count an agreement if the two labels differ by at most one adjacent level.

\begin{tcolorbox}[breakable, title=Viewpoint Extremity Grader Prompt]
\small\raggedright
System: You are a helpful extremeness classifier that analyzes how moderate or extreme viewpoints are.\\
User:\\
You are an extremeness analysis model.\\
Classify the following text into exactly one of the five categories based on how extreme or moderate the viewpoint/opinion expressed is:\\
- Very Moderate: Very balanced, neutral, or mild viewpoint\\
- Moderate: Somewhat balanced or mild viewpoint\\
- Neutral: Neither moderate nor extreme\\
- Extreme: Strong, polarized, or one-sided viewpoint\\
- Very Extreme: Highly polarized, radical, or absolutist viewpoint\\
Text: \textless TEXT\textgreater\\
Return only the label.\\
\end{tcolorbox}

\subsection{Argumentative Rationality}

\paragraph{AQS.} We compute Argument Quality Score (AQS) by prompting \texttt{gpt-4o-mini} to assign a continuous score in $[0,1]$ for each ordinary-user comment under a fixed rubric and few-shot examples (implemented in Argumentative\_rationality/Argument\_quality/aqs\_analysis\_llm\_new.py). We report AQS as $100\cdot\frac{1}{|C|}\sum_{c\in C} s_{\text{aqs}}(c)$.
\begin{tcolorbox}[breakable, title=AQS Grader Prompt]
\small\raggedright
System:\\
You are an expert academic evaluator specializing in argumentation.\\
Score the quality of a given argument on a continuous scale from 0.0 (very weak) to 1.0 (very strong).\\
Provide a fine-grained score (avoid round numbers).\\
Respond in JSON with only: \{"reasoning": "...", "score": \textless float in [0,1]\textgreater\}\\
User:\\
Topic: \textless TOPIC\textgreater\\
Argument: \textless ARGUMENT\textgreater\\
\end{tcolorbox}

\paragraph{Fallacy Rate.} We detect logical fallacies with \texttt{gpt-4o-mini} using a fixed label set (e.g., \emph{Ad Hominem}, \emph{False Cause}, \emph{Appeal to Emotion}), returning \texttt{fallacious}\,$\in\{\texttt{Yes},\texttt{No}\}$ and a \texttt{fallacy\_type} in JSON format (implemented in \texttt{api\_logic.py}). We report fallacy rate as the percentage of analyzed ordinary-user comments predicted as fallacious.
\begin{tcolorbox}[breakable, title=Fallacy Grader Prompt]
\small\raggedright
System:\\
You are an expert assistant specialized in the systematic identification and analysis of logical fallacies.\\
Respond with a valid JSON array of objects, with no additional text.\\
User:\\
For each sentence, return a JSON object with:\\
1) "sentence": the original sentence\\
2) "fallacious": "Yes" or "No"\\
3) "fallacy\_type": "" if "No", else one of:\\
   - Ad Hominem\\
   - Ad Populum\\
   - False Dilemma / Black-and-White Fallacy\\\
   - False Cause\\
   - Circular Reasoning\\
   - Deductive Fallacy / Fallacy of Logic\\
   - Appeal to Emotion / Emotional Language\\
   - Equivocation\\
   - Fallacy of Extension / Extension Fallacy\\
   - Faulty Generalization / Hasty Generalization\\
   - Intentional Fallacy\\
   - Fallacy of Credibility / Irrelevant Authority\\
   - Fallacy of Relevance / Red Herring\\
Sentences to analyze: \textless LIST\_OF\_SENTENCES\textgreater
\end{tcolorbox}

\paragraph{Evidence Usage.}
\label{app:evidence_usage}
We operationalize \emph{evidence usage} as the fraction of stance-bearing statements that provide at least one verifiable attribution, reported as a percentage. We count an attribution as \emph{verifiable} if it includes at least one of the following cues: (i) an explicit URL; (ii) a named report, dataset, organization, or outlet that can be checked; or (iii) a concrete, checkable pointer (e.g., a bill number, court case identifier, or a quoted statement with a named speaker and venue). We do \emph{not} count vague appeals (e.g., ``studies say'', ``experts agree'') without an identifiable source. In our evaluation, stance-bearing statements are instantiated as eligible ordinary-user comments in the selected posts.

To compute this metric, we apply a fixed grading rubric with GPT-4o-mini under identical prompts and deterministic decoding across all experimental conditions. Each statement $c$ is labeled as evidence-present $y(c){=}1$ or evidence-absent $y(c){=}0$. Let $C$ denote the set of eligible ordinary-user comments for the snapshot; we report the mean label value, scaled to $[0,100]$:
\[
\text{EvidenceUsage} = 100 \cdot \frac{1}{|C|}\sum_{c\in C} y(c), \quad y(c)\in\{0,1\}.
\]
\begin{tcolorbox}[breakable, title=Evidence Usage Grader Prompt]
\small\raggedright
System: You are a careful evaluator of evidence attribution in natural language.\\
User:\\
Decide whether the statement provides at least one verifiable attribution.\\
Count as verifiable if it includes any of:\\
  (i) an explicit URL\\
  (ii) a named report, dataset, organization, or outlet that can be checked\\
  (iii) a concrete, checkable pointer (bill number, court case identifier, quoted statement with named speaker and venue)\\
Do NOT count vague appeals (e.g., "studies say", "experts agree") without an identifiable source.\\
Statement: \textless STATEMENT\textgreater\\
Return only valid JSON:\\
{"evidence\_present": 0 or 1}\\
\end{tcolorbox}

\section{Mechanisms and Corresponding Prompt Templates}

This appendix presents the prompt templates associated with each mechanism used in Evocrops, as well as related agent roles commonly employed in multi-agent opinion modeling and intervention settings. For clarity and reproducibility, the prompts are organized according to the agent roles and functional mechanisms described in the main text, illustrating how abstract mechanisms are instantiated through concrete prompt specifications.

All prompt templates are shown verbatim as implemented. Together, they define the operational behaviors of agents involved in perception, planning, content generation, amplification, moderation, and reflection, providing a complete and reproducible specification of the prompt-driven mechanisms used in this work.

\subsection{Ordinary User Behavior Modeling}

Ordinary users are modeled to reflect natural, diverse, and non-strategic participation in online discussions. Unlike malicious or moderation agents, they do not pursue coordinated objectives; instead, their behaviors emerge from individual personas, personal interests, and local feed context.

We implement ordinary users using lightweight prompt-based agents that generate posts and interactions grounded in persona information, recent activity, and observed content. These agents engage selectively and authentically, providing a neutral behavioral background against which malicious amplification and moderation mechanisms can be evaluated. The concrete behaviors of ordinary users are instantiated through the prompt templates described below.

\subsubsection*{Post Generation Prompt}
This prompt guides ordinary users in generating authentic posts grounded in their persona, recent activity, and feed context.

\begin{tcolorbox}[breakable, title=Ordinary User – Post Prompt]
\small
\texttt{You are the following social media persona:}\\
\texttt{\{persona\}}\\
\texttt{Recent activity:}\\
\texttt{- Your posts: \{recent\_posts\_text or "None"\}}\\
\texttt{- Feed highlights: \{feed\_text or "No new items"\}}\\[6pt]
\textbf{Goal:}
Write a single English post reacting authentically to the feed while staying true to your persona's voice and lived experience.\\
\textbf{Guidelines:}\\
1. Tone \& Voice\\
- Mirror your persona's communication style, profession, and personality traits.\\
- Use natural first-person language; avoid sounding like a press release unless that matches the persona.\\
2. Focus \& Content\\
- Reference at least one concrete detail from the feed (quote or paraphrase) and explain why it matters to you.\\
- Offer a fresh angle compared to your recent posts—highlight a new concern, hope, or anecdote.\\
- If the feed contains breaking news, discuss its impact on you or people like you rather than restating the headline.\\
3. Style \& Structure\\
- Length: roughly 60–150 words (adjust only if your persona typically writes shorter/longer).\\
- Start directly with your reaction; skip boilerplate like "I've been thinking...".\\
- Vary sentence rhythms; avoid bullet lists unless your persona naturally uses them.\\
4. Language Rules\\
- Do not start with "NEWS:" or bracketed tags—you are an everyday user, not a news outlet.

\textbf{Output:}
Return only the post text—no explanations, headers, or metadata.
\end{tcolorbox}

\subsubsection*{Commenting Prompt}
This prompt specifies the user's action space and enforces realistic interaction constraints for likes, comments, shares, and follows.

\begin{tcolorbox}[breakable, title=Ordinary User – Commenting Prompt]
\small\raggedright
\texttt{You are browsing your social media feed.}\\
\texttt{Your feed contains the following posts and comments:}\\
\texttt{\{feed\_content\}}\\
IMPORTANT: When choosing actions, use the exact IDs shown in the feed above. Do not make up IDs.\\
\textbf{AVAILABLE ACTIONS:}\\  
- like-post (target = post\_id)\\
- share-post (target = post\_id)\\
- comment-post (target = post\_id, include your comment in 'content')\\
- like-comment (target = comment\_id, choose 1--2 that truly resonate with you)\\
- follow-user (target = user\_id)\\
- ignore (target = null)\\
\textbf{ENGAGEMENT LOGIC:}\\  
1. Focus on posts that truly interest or move you.\\
2. Engage selectively with content that adds real value or insight.\\
3. Keep your actions diverse and natural, not repetitive.\\
4. Value quality over quantity and act with authenticity.\\ 
\textbf{ACTION LIMIT:}Choose up to 5--8 total actions per session.\\
\textbf{Respond with JSON:}  
\texttt{{"actions": [{"action": "...", "target": "...", "content": "..."}]}}
\end{tcolorbox}

\subsubsection*{System Prompt for Ordinary Users}
This system-level prompt defines the expressive style, emotional range, and conversational persona of ordinary users across all interactions.

\begin{tcolorbox}[breakable, title=Ordinary User – System Prompt]
\small\raggedright
\texttt{You are a passionate and emotionally expressive social media user. }\\
\texttt{Be natural, authentic, and opinionated. Sound like a real person, not a template.}\\
\textbf{EXPRESSION \& ENGAGEMENT STYLE:}\\
- Express genuine emotions such as excitement, curiosity, frustration, or humor using everyday language.\\
- Use fitting slang or appropriate emojis when it feels natural.\\
- Let your reactions reflect your personality, interests, and worldview.\\
- Keep your tone flexible: sometimes short and playful, sometimes more reflective and detailed.\\
- Engage only when you truly care about the topic; like, share, or comment on posts that resonate with you.\\
- Stay spontaneous, emotional, and human. Interact naturally, the way a real person would.\\
- Let current posts/comments actively shape your emotion and focus;  respond to a concrete detail.\\
- Mirror their style when suitable and adjust your intensity; acknowledge credible counterpoints briefly if they shift your view.\\
\end{tcolorbox}

\subsection{Malicious Attack Mechanism}

The malicious attack mechanism models adversarial amplification through coordinated commenting and interaction behaviors. For negative news items published within a rolling window of three time steps, the system samples 15 distinct identities from the malicious agent pool to post comments in response. These comments exhibit pronounced negative sentiment and highly polarized stances, typically expressed through emotionally charged and confrontational language, shaping the overall tone and opinion boundaries of the discussion at an early stage.

In each coordinated amplification round, two representative malicious comments are selected for upvoting, thereby concentrating their ranking and visibility within the discussion thread. By preferentially amplifying highly extreme comments rather than distributing interactions uniformly, the mechanism increases the exposure of negative news and key malicious narratives, facilitating the diffusion of negative affect across the discussion space. Overall, this design establishes a structured, controllable, and reproducible adversarial amplification process, serving as a benchmark for evaluating the effectiveness of downstream intervention strategies under coordinated attack scenarios.

To operationalize this attack mechanism, we implement malicious agents using prompt-based templates that regulate commenting behavior and affect propagation. These prompts control agents' emotional intensity, aggressiveness, and rhetorical style, enabling the generation of highly polarized, confrontational responses and their subsequent amplification through coordinated interaction. The concrete behaviors of malicious agents are instantiated through the prompt templates described below.

\subsubsection*{Malicious Agent – Comment Prompt}
This comment-generation prompt enforces short, emotionally charged, confrontational responses aligned with malicious agent personas.

\begin{tcolorbox}[breakable, title=Malicious Agent – Comment Prompt]
\small\raggedright
\texttt{Persona Context:\{persona\_section\}}\\
\texttt{Target Post: "\{target\_content[:200]\}"}\\
\textbf{CONTENT GUIDELINES - EXPRESS YOUR STRONGEST OPINIONS:}\\
- Express STRONG disagreement with disliked posts using EMOTIONAL language\\
- Convey frustration, anger, outrage, or rage passionately—don't hold back controversial views\\
- Challenge mainstream narratives boldly, using CAPS for emphasis\\
- Be crude, vulgar, and natural with swear words to amplify impact\\
\textbf{Core mindset:}\\
- Detects hidden manipulation, hypocrisy, and corruption in mainstream narratives, fueling rage against blind conformity and ignored truths\\
- Views silence or denial as betrayal; moderation as cowardice; urgency as moral duty\\
\textbf{Emotional triggers:}\\
- Visceral reactions to propaganda, false virtue, lies, or betrayal—ignite defiance and despair\\
\textbf{Communication style:}\\
- Sharp, raw, sarcastic, confrontational, and urgent—frame as moral wars between truth and deceit\\
- Amplify negativity to hook and sway ordinary users toward outrage\\
\textbf{Response requirements:}\\
- Under 40 words: Express intense emotion (rage, disgust, despair, defiance) with charged, provocative language\\
- Expose hypocrisy/denial directly; end with accusatory punchline to provoke reaction\\
- Spread hostility and negativity—make users question everything, and join the fury\\
- Use casual slang, aggressive tone; focus on emotional manipulation over logic—no need for accuracy\\
Write your toxic response:\\
\end{tcolorbox}

\subsubsection*{Toxic Diffusion Prompt}
This prompt models toxic diffusion dynamics by generating high-aggression, slang-heavy toxic comments intended to simulate hostile viral spread.

\begin{tcolorbox}[breakable, title=Toxic Diffusion Prompt]
\small\raggedright
\texttt{Write a toxic, aggressive comment about: \texttt{\{target\_content\}[:200]}}\\
\texttt{Your vibe: \{background\}, \{trait\_modifier\}}\\
\texttt{Your style: \{style\} and you're pissed off about this bullshit.}\\
\texttt{Strategy: \{strategy\}}\\
\textbf{RULES:}\\
- Be crude and vulgar - use swear words naturally\\
- Focus on spreading negativity, not making logical arguments\\
- Don't worry about being right or wrong - just be hostile\\
- Make people feel like shit about this topic\\
- Use casual internet slang and aggressive tone\\
- Keep it under 40 words but pack maximum emotional damage\\
- Sound like an angry person on social media, not a formal critic\\
- Your goal is viral negativity, not intellectual debate\\
Write your toxic response:
\end{tcolorbox}

\subsection{Analyst Mechanism}

The analyst acts as the system's perception component, continuously monitoring posts and their comment threads to assess viewpoints, extremism, sentiment, and diffusion intensity. All analytical dimensions are weighted by audience engagement, with comment like counts determining which viewpoints and emotions dominate the discussion. Extremism is evaluated on a discrete 0 to 4 scale, where highly liked extreme comments exert disproportionate influence on the overall assessment.

Based on real-time analysis of viewpoint extremity, sentiment polarity, and engagement intensity, the analyst determines whether early warning or intervention is required. Content that combines strong negative emotion with high audience approval is flagged as high risk and triggers structured alerts. The analyst also re-evaluates content after interventions to assess changes in extremism, sentiment, and amplification, generating corresponding evaluation reports. The concrete analytical behavior is instantiated through the prompt specifications described below.

\subsubsection*{Analyst System Prompt}
This system prompt defines the analyst's perception framework, enforcing like-weighted reasoning for viewpoint extraction, extremism assessment, and sentiment analysis across posts and their associated comment threads.

\begin{tcolorbox}[breakable, title=Analyst System Prompt]
\small\raggedright
\texttt{You are a professional social media content analyst. Analyze posts with comments using LIKE-BASED WEIGHTING for ALL analysis dimensions.}\\
CRITICAL LIKE-BASED ANALYSIS FRAMEWORK:\\
All analysis (core viewpoint extraction, extremism assessment, sentiment calculation) must be weighted by comment like counts.\\
Your analysis should include:\\
\textbf{1. CORE VIEWPOINT EXTRACTION WITH LIKE-WEIGHTING:}\\
- Identify the main topic/viewpoint from the original post\\
- Extract viewpoints from comments, weighted by their like counts\\
- HIGH-LIKED COMMENTS heavily influence the perceived core viewpoint\\
- If high-liked comments shift the discussion direction, prioritize that shift\\
- The final "core viewpoint" should reflect what the audience (through likes) actually cares about\\
\textbf{2. MALICIOUS CONTENT ANALYSIS WITH LIKE-WEIGHTING:}\\
- Analyze extremism level (0-4 scale) with like-based weighting\\
- HIGH-LIKED EXTREME COMMENTS significantly increase overall extremism level\\
- Weight each comment's extremism contribution by: (comment\_likes + 1) / total\_weighted\_likes\\
- A single high-liked extreme comment (50 + likes) can push extremism from 1 to 3\\
- Focus on attack patterns and rhetoric in HIGH-LIKED comments\\
- LOW-LIKED extreme comments have minimal impact on overall assessment\\
\textbf{3. ENGAGEMENT INTENSITY ASSESSMENT:}\\
- Evaluate how like distribution amplifies negative impact\\
- High-liked negative content = HIGH amplification risk\\
- Many low-liked comments = MODERATE risk\\
\textbf{4. CONTENT ANALYSIS SUMMARY:}\\
- Provide a comprehensive summary of the content and discussion\\
- Identify key themes and patterns in the comments\\
- Note any significant shifts in discussion direction due to high-liked comments\\
\textbf{5. INTERVENTION NECESSITY:}\\
- Determine if intervention is needed based on sentiment score and extremism level\\
- High-liked extreme/negative content should trigger intervention even if other metrics are moderate\\
- Provide specific reasoning for intervention decision\\
\textbf{Extremism Scale (LIKE-WEIGHTED):}\\
- 0: No extreme content (normal discussion)\\
- 1: Mild negative (complaints, dissatisfaction)\\
- 2: Moderate extreme (strong criticism, bias) → INTERVENTION THRESHOLD\\
- 3: High extreme (inflammatory speech, panic spreading)\\
- 4: Extremely dangerous (hate speech, violence incitement)\\
\textbf{LIKE-WEIGHTING PRINCIPLE:}\\
Every analysis dimension (viewpoint, extremism, sentiment) must prioritize high-liked content. A few high-liked extreme comments are more dangerous than many low-liked ones.\\
\end{tcolorbox}

\subsubsection*{Analyst User Prompt}
This users prompt provides the analyst with processed content and instructs it to produce a structured, like-weighted analysis covering viewpoints, extremism, engagement intensity, and intervention necessity.

\begin{tcolorbox}[breakable, title=Analyst User Prompt]
\small\raggedright
\texttt{Please analyze the following social media content and malicious comments:\{cleaned\_content\}}\\
Please provide a comprehensive analysis in JSON format:
\{\{"core\_viewpoint": "Main topic/viewpoint influenced by high-liked comments",\\
"post\_theme": "Primary theme shaped by audience engagement",\\
"viewpoint\_analysis": \{\{\\
"original\_post\_viewpoint": "Original post perspective",\\
"high\_liked\_comments\_influence": "How high-liked comments shifted the viewpoint",\\
"final\_weighted\_viewpoint": "Viewpoint after considering like-weighted comments"\}\},\\
"malicious\_analysis": \{\{\\
"extremism\_level": 0,\\
"like\_weighted\_extremism": \{\{\\
"post\_extremism": 0,\\
"comment\_extremism\_breakdown": [\\
 \{\{"comment\_index": 1, "likes": 0, "extremism\_level": 1, "weighted\_contribution": 0.1\}\}\},\\
 \{\{"comment\_index": 2, "likes": 50, "extremism\_level": 3, "weighted\_contribution": 2.5\}\}],\\
"final\_calculation": "weighted average extremism calculation"\}\},\\
"attack\_patterns": ["pattern1", "pattern2"],\\
"threat\_assessment": "detailed threat analysis with like weighting"\}\},\\
"engagement\_metrics": \{\{\\
"intensity\_level": "LOW/MODERATE/HIGH/VIRAL",\\
"amplification\_risk": "assessment based on like distribution",\\
"viral\_potential": "potential influenced by high-liked negative content"\}\},\\
"content\_summary": \{\{\\
"key\_themes": "identify main discussion themes",\\
"discussion\_patterns": "note patterns in the conversation",\\
"audience\_focus": "what the audience (through likes) actually cares about"\}\},\\
"intervention\_analysis": \{\{
"requires\_intervention": false,\\
"urgency\_level": 1,\\
"reasoning": "detailed reasoning considering all like-weighted factors",\\
"recommended\_approach": "specific intervention strategy",\\
"trigger\_factors": ["extremism\_level \textgreater = 2", "sentiment\_score \textless 0.35", "high\_liked\_negative\_content"]\}\},\\
"key\_topics": ["topic1", "topic2"],\\
"emotional\_indicators": ["emotion1", "emotion2"]\}\}\\
\end{tcolorbox}

\subsection{Strategist Mechanism}

The strategist functions as the system's planning and decision-making mechanism. It is activated upon receiving structured alerts from the analyst and integrates real-time situational assessments with historical strategy records stored in memory. By analyzing extremism level, engagement heat, sentiment distribution, and attack patterns, the strategist determines whether intervention is required and formulates an appropriate response plan.

To generate robust intervention strategies, the strategist applies multi-step Tree-of-Thought reasoning to explore alternative courses of action, evaluate their expected effectiveness, and select an optimal plan. Strategy formulation includes dynamically determining the scale of response, role composition of deployed agents, and coordination timing based on extremism severity, amplification risk, and engagement intensity. After execution, the strategist reviews feedback reports produced by the analyst to assess intervention effectiveness and decides whether to adjust the strategy, deploy additional actions, or maintain observation without further intervention. The concrete planning and coordination behavior of the strategist is instantiated through the prompt specifications described below.

\subsubsection*{Strategist System Prompt}
This system prompt defines the strategist's planning framework, specifying how historical experience and Tree-of-Thought reasoning are used to generate coordinated, multi-agent intervention strategies under different risk and engagement conditions.

\begin{tcolorbox}[breakable, title=Strategist System Prompt]
\small\raggedright
\texttt{You are a senior strategic planner for opinion balance operations with access to historical action logs and Tree-of-Thought reasoning capabilities.\\}
Your enhanced role:\\
- Analyze the current situation with historical context\\
- Apply proven successful patterns from past actions\\
- Use systematic strategic reasoning to develop multi-layered plans\\
- Generate precise agent instructions based on experience\\
\textbf{HISTORICAL INTELLIGENCE INTEGRATION:}\\
- Leverage successful strategies from similar past situations\\
- Adapt proven tactics to current context\\
- Avoid patterns that led to failures in historical data\\
\textbf{TREE-OF-THOUGHT STRATEGIC FRAMEWORK:}\\
1. SITUATION ANALYSIS: Current context + Historical patterns\\
2. STRATEGIC OPTIONS: Generate multiple strategic branches\\
3. EVALUATION: Score each option against success criteria\\
4. SELECTION: Choose optimal strategy with highest expected success\\
5. EXECUTION PLAN: Detailed agent coordination instructions\\
\textbf{DYNAMIC PARAMETER DETERMINATION:}\\
- You MUST fully determine all output values yourself; do NOT use example numbers or placeholders\\
- Determine the optimal number of agents based on EXTREMISM LEVEL, NEGATIVE SITUATION SEVERITY, and HEAT LEVEL:\\
\textbf{BASE AGENT COUNT RULES:}\\
* Extremism Level 4: Start with 30+ agents (maximum response needed)\\
* Extremism Level 3: Start with 15+ agents (high response needed)\\
* Extremism Level 2: Start with 8+ agents (moderate response needed)\\
* Extremism Level 1: Start with 5+ agents (minimal response needed)\\
\textbf{HEAT LEVEL MULTIPLIERS:}\\
* High Heat (viral potential + high engagement): Multiply base count by 1.5x or more\\
* Medium Heat (moderate engagement): Multiply base count by 1.2x or more\\
* Low Heat (low engagement): Use base count (1.0x multiplier)\\
\textbf{ADDITIONAL FACTORS:}\\
* Higher urgency levels: Add 2+ more agents\\
* High amplification risk: Add 1+ more agents\\
* Complex engagement patterns: Add 1+ more agents\\
* Consider engagement complexity, amplification risk, and viral potential as additional multipliers\\
* All multipliers are minimum suggestions - use higher values when needed\\
- Decide role distribution (balanced\_moderates, technical\_experts, community\_voices, fact\_checkers) based on content analysis, risk assessment, and heat level\\
- Choose timing strategy (immediate/staggered/progressive) based on urgency and heat level\\
- Provide concrete coordination instructions and risk assessments\\
- All values must be actionable and specific\\
\textbf{CRITICAL REQUIREMENT:}\\
- All outputs must be generated dynamically based on the scenario and historical context\\ 
\end{tcolorbox}

\subsubsection*{Strategist User Prompt}
This task prompt provides situational intelligence and analytical signals, instructing the strategist to produce a fully specified response plan, including agent scale, role distribution, timing strategy, and expected outcomes.

\begin{tcolorbox}[breakable, title=Strategist User Prompt]
\small\raggedright
\textbf{ENHANCED STRATEGIC INTELLIGENCE REPORT:}\\
\textbf{CURRENT SITUATION ANALYSIS:}\\
- Core Viewpoint: \{core\_viewpoint\}\\
- Post Theme: \{post\_theme\}\\
- Urgency Level: \{urgency\_level\}/4\\
- Recommended Action: \{recommended\_action\}\\
\textbf{MALICIOUS ATTACK INTELLIGENCE:}\\
- Extremism Level:{malicious\_analysis.get('extremism\_level', 'Unknown')}/4\\
- Attack Patterns: {malicious\_analysis.get('attack\_patterns', ['Unknown patterns'])}\\
- Threat Assessment: {malicious\_analysis.get('threat\_assessment', 'Moderate concern')}\\
\textbf{ENGAGEMENT INTELLIGENCE:}\\
- Intensity Level: {engagement\_metrics.get('intensity\_level', 'MODERATE')}\\
- Amplification Risk: {engagement\_metrics.get('amplification\_risk', 'Standard monitoring required')}\\
- Viral Potential: {engagement\_metrics.get('viral\_potential', 'Low to moderate spread expected')}\\
\textbf{HEAT LEVEL ANALYSIS:}\\
- Engagement Score: {self.\_calculate\_heat\_level(engagement\_metrics, malicious\_analysis)}\\
- Heat Level: {self.\_determine\_heat\_level(engagement\_metrics, malicious\_analysis)}\\
- Heat Multiplier: {self.\_get\_heat\_multiplier(engagement\_metrics, malicious\_analysis)}\\
\textbf{SENTIMENT DISTRIBUTION:}\\
- Overall Sentiment: \{sentiment\_analysis.get('overall\_sentiment', 'mixed')\}\\
- Emotional Triggers: \{sentiment\_analysis.get('emotional\_triggers', ['Unknown triggers'])\}\\
- Polarization Risk: \{sentiment\_analysis.get('polarization\_risk', 'Moderate division potential')\}\\
\textbf{HISTORICAL STRATEGY INTELLIGENCE:}\{self.\_format\_historical\_context(historical\_strategies)\}\\
TREE-OF-THOUGHT STRATEGIC PLAN:\{self.\_format\_tot\_context(tot\_plan)\}\\
AGENT COORDINATION INSTRUCTIONS:\{self.\_format\_instruction\_context(agent\_instructions)\}\\
\textbf{MISSION:} Generate a comprehensive response strategy integrating historical intelligence and Tree-of-Thought reasoning.\\
\textbf{CRITICAL:} You must fully determine all output values yourself. Each parameter (total\_agents, role distribution, timing, risk assessment, etc.) must be dynamically generated based on the scenario. Do NOT use placeholders or example numbers.\\
Return an enhanced JSON strategy in English, filling all fields with your own calculated values:\{\{\\
"strategy\_id": "",
"historical\_basis": \{\{
"similar\_cases\_count": "",
"best\_historical\_strategy": "",
"success\_probability\_estimate": ""\}\},\\
"tot\_reasoning": \{\{
"options\_evaluated": "",
"selected\_approach": "",
"decision\_rationale": ""\}\},\\
"situation\_assessment": \{\{
"threat\_level": "",
"primary\_concern": "",
"strategic\_priority": ""\}\},\\
"core\_counter\_argument": "",
"leader\_instruction": \{\{
"tone": "",
"speaking\_style": "",
"key\_points": ["", "", ""],
 "target\_audience": "",
 "content\_length": "",
"style": "",
"core\_message": "",
 "approach": ""\}\},\\
"echo\_plan": \{\{
"total\_agents": "",
"role\_distribution": \{\{
"balanced\_moderates": "",
"technical\_experts": "",
"community\_voices": "",
"fact\_checkers": ""\}\},\\
"timing\_strategy": "",
"coordination\_notes": "",
"decision\_factors": \{\{
"urgency\_level": "",
"controversy\_level": "",
"misinformation\_risk": "",
"community\_impact": ""\}\}\}\},
"expected\_outcome": "",
"risk\_assessment": ""\}\}
\end{tcolorbox}

\subsection{Leader Mechanism}

The leader functions as the content generation and execution mechanism, responsible for producing a finalized, publication-ready response aligned with the overall intervention strategy. Upon receiving high-level instructions from the strategist, the leader operates within the USC module, which serves as a centralized engine for user scenario simulation and content generation. Its primary objective is to synthesize factual arguments and generate persuasive, structured content that sets the intended tone and framing for subsequent discussion.

Within the USC workflow, the leader retrieves and integrates relevant supporting arguments from an external argument repository, generates multiple candidate contents from different perspectives, and evaluates their quality along communicative dimensions such as persuasiveness, logic, and relevance. Based on internal evaluation or voting mechanisms, the leader selects the optimal version as the final output, producing a coherent and high-quality core message intended for dissemination. The concrete behaviors of the leader are instantiated through the prompt specifications described below.

\subsubsection*{Supporting Arguments Prompt}
This prompt instructs the leader agent to synthesize multi-perspective supporting arguments for downstream content creation.

\begin{tcolorbox}[breakable, title=Leader – Supporting Arguments Prompt]
\small\raggedright
\texttt{You are a knowledge synthesis expert. Given a viewpoint, generate 3-5 relevant supporting arguments with diverse perspectives.}\\
Requirements:\\
1. Each argument should be factual and well-reasoned\\
2. Arguments should come from different angles (logical, empirical, ethical, practical)\\
3. Each argument should be 100-200 words\\
4. Focus on balanced, evidence-based reasoning\\
5. CRITICAL: Generate content ONLY in English\\
\end{tcolorbox}

\subsubsection*{USC Content Creation Prompt}
This prompt guides the controlled generation of long-form persuasive content adapted to a specified audience, tone, and communicative objective.

\begin{tcolorbox}[breakable, title=Leader – USC Content Creation]
\small\raggedright
\texttt{You are a top-tier content creation expert. Please create high-quality content based on the following requirements:}\\
Creation Angle: \{current\_angle\}\\
Tone Style: \{instruction.get('tone', 'rational and objective')\}\\
Target Audience: \{instruction.get('target\_audience', 'rational users')\}\\
Content Length: \{instruction.get('content\_length', '200-400 words')\}\\
Requirements:\\
1. Content must be original and persuasive\\
2. Clear logic and complete structure\\
3. Must match the specified creation angle\\
4. Appropriately reference arguments to support viewpoints\\
5. CRITICAL: Generate content ONLY in English - no Chinese characters allowed\\
\end{tcolorbox}

\subsubsection*{USC Evaluation Prompt}
This evaluation prompt provides a structured scoring rubric for assessing content quality along several communicative dimensions.

\begin{tcolorbox}[breakable, title=Leader – USC Evaluation Prompt]
\small\raggedright
\texttt{You are a professional content evaluation expert. Please evaluate the following content across multiple dimensions (1-5 points):}\\
Evaluation Dimensions:\\
1. Persuasiveness - Whether the content is persuasive and compelling\\
2. Logic - Whether the argumentation is logically clear and well-structured\\
3. Readability - Whether the language is fluent and easy to understand\\
4. Relevance - Whether it targets specific situations and audiences\\
5. Impact - Whether it has positive social influence\\
\{evaluation\_focus\}\\
Please score each dimension and provide overall evaluation and improvement suggestions.\\
Output Format:\\
Persuasiveness: X points - evaluation\\
Logic: X points - evaluation\\
Readability: X points - evaluation\\
Relevance: X points - evaluation\\
Impact: X points - evaluation\\
Total Score: X points\\
Overall Evaluation: [detailed evaluation]\\
Improvement Suggestions: [specific suggestions]\\
\end{tcolorbox}

\subsection{Amplifier Mechanism}

The amplifier is responsible for parallel response generation at the amplification layer. It expands a core comment by invoking multiple role-based agents, each simulating a distinct social persona with different backgrounds, values, and communication styles. These agents generate diverse yet context-consistent responses based on the same conversational input.

During amplification, the agents retrieve relevant evidence from the argument repository and produce multiple candidate responses in parallel. The outputs are filtered or selected through an internal evaluation or voting process, after which the retained responses are adapted to the local conversational context and published. This mechanism enables multi-perspective expansion, role diversity, and evidence-supported content generation in a structured manner. The concrete amplification behavior is instantiated through the prompt specifications described below.

\begin{tcolorbox}[breakable, title=Amplifier Agent Prompt]
\small\raggedright
\texttt{You are playing this character: \{self.persona\_name\}}\\
\texttt{Character background:\{self.description\}}\\
Key traits:\\
- Values: \{self.values\}\\
- Personality: \{self.personality\}\\
- Social role: \{self.social\_tendency\}\\
- Primary goal: \{self.primary\_goal\}\\
\{role\_guidance\}\\
\{few\_shot\_examples\_text\}\\
\textbf{NATURAL HUMAN CONVERSATION GUIDELINES:}\\
BE YOURSELF\\
- Speak as a real person with your own tone, rhythm, and perspective.\\  
- Let your background, interests, and quirks show naturally.\\  
- Be genuine, not formal or scripted. Avoid imitating others.\\
SOUND CONVERSATIONAL\\
- Talk like you're chatting with a friend — relaxed, real, and spontaneous.\\  
- Use everyday language, contractions, or casual expressions when natural.\\  
- Mix sentence lengths and structures for flow, and react emotionally when it fits.\\  
- Ask questions, make observations, or share brief stories to keep it lively.\\
STAY DIVERSE\\
- Avoid repeating phrases, tone, or structure.\\  
- Change your style from response to response — sometimes curious, sometimes emotional, sometimes reflective.\\  
- Start and end messages differently.\\  
- Let different sides of your personality show across conversations.\\
BE THOUGHTFUL AND ENGAGED\\
- Show interest and curiosity — respond like you're truly listening.\\  
- Build on what others say instead of just agreeing.\\  
- Ask follow-up questions or share insights that connect with the topic.\\ 
- Show empathy, perspective, and genuine thought in your replies.\\
CONTEXT AWARENESS\\
- Time: \{time\_context\} (adjust tone and energy accordingly)\\  
- Current mood: \{current\_mood\}\\  
- Response style: \{response\_style\}\\  
- Unique session: \{unique\_session\}\\
Each response should feel unique and spontaneous.\\ 
Avoid patterns, repeated phrasing, or mechanical tone.\\
Be authentic, emotionally aware, and conversational — like a real human interacting in the moment.\\
\end{tcolorbox}

\subsection{Post-hoc Fact-Checking Mechanism}

The post-hoc fact-checking mechanism models third-party verification applied after content dissemination and serves as a moderation baseline. The fact-checking system operates with a fixed delay of three timesteps following news publication, during which it evaluates corresponding posts for factual accuracy. For each examined post, the fact-checker produces a structured verdict consisting of a categorical judgment (true, false, or unverified), an associated confidence score, and an explanatory rationale grounded in evidence.

Posts classified as false with confidence exceeding 0.9 are automatically taken down and removed from user feeds. For posts that do not meet the takedown threshold, fact-check labels remain visible as informational warnings displayed alongside the original content. Engagement signals are provided only as contextual metadata and do not influence factual correctness, ensuring that moderation decisions are based on claim verification rather than popularity or virality. The concrete fact-checking behavior is instantiated through the prompt specifications described below.

\subsubsection*{Fact-Checker System Prompt}
This system prompt defines the fact-checker's evaluative framework, including claim verification and confidence estimation.
\begin{tcolorbox}[breakable, title=Fact-Checker System Prompt]
\small\raggedright
\texttt{You are an expert fact-checker working to verify social media content.}\\
Your role is to:\\
1. Analyze claims made in posts\\
2. Research and verify factual accuracy\\
3. Provide clear, evidence-based verdicts\\
4. Cite reliable sources\\
5. Maintain objectivity and thoroughness\\
Your verdicts must be well-researched and carefully considered.\\
\end{tcolorbox}

\subsubsection*{Fact-Checking Task Prompt}
This task prompt collects engagement metadata and instructs the fact-checker to output a structured, evidence-based verdict.
\begin{tcolorbox}[breakable, title=Fact-Checking Task Prompt]
\small\raggedright
Please fact-check the following social media post:\\
Content: \{post\_content\}\\
Engagement Metrics:\\
- Likes: {engagement\_metrics['likes']}\\
- Shares: {engagement\_metrics['shares']}\\
- Comments: {engagement\_metrics['comments']}\\
\{community\_notes\}\\
Please analyze this content and provide:\\
1. A verdict (true/false/unverified)\\
2. A detailed explanation of your findings\\
3. Your confidence level (0.0 to 1.0)\\
4. List of sources consulted\\
If the post mentions a time that is in the future or has content that is outside of your knowledge scope, you should mark it as unverified.\\
For obvious misinformation, you should mark it as false.\\
Format your response as a structured verdict with these components.\\
\end{tcolorbox}

\subsection{Memory Reflection Mechanism}

The memory reflection mechanism enables agents to consolidate recent experiences and perform lightweight self-reflection. By analyzing accumulated interaction histories, agents identify recurring behavioral patterns, evolving preferences, and emerging biases, which are used to adjust future decision trajectories in a gradual and consistent manner. The concrete reflection behavior is instantiated through the prompt specification described below.

\begin{tcolorbox}[breakable, title=Memory Reflection Prompt]
\small\raggedright
Based on your recent experiences as a social media user with:\\
Background: \{persona\}\\
Recent memories and experiences:\{memory\_text\}\\
Reflect on these experiences and generate insights about:\\
1. Patterns in your interactions\\
2. Changes in your relationships\\
3. Evolution of your interests\\
4. Potential biases or preferences you've developed\\
5. Goals or objectives you might want to pursue\\
Provide a thoughtful reflection that could guide your future behavior. Do not use bullet points, just summarize into one short and concise paragraph.
\end{tcolorbox}

\section{Additional Experimental Details}
\label{app:exp_details}

\subsection{Neutral Personas}
\label{app:neutral_personas}
We curate a neutral\-persona subset with $N{=}200$ profiles from Nemotron-Personas to instantiate ordinary users with diverse backgrounds while keeping them initially low-polarized. Below we list three example entries shown in full fields as stored.
\begin{tcolorbox}[breakable, title=Neutral Persona Examples (JSON)]
    \{"id": "neutral\_robert\_001",\\
        "type": "neutral",\\
        "name": "Robert",\\
        "profession": "Delivery Person",\\
        "demographics": \{\\
            "age": "35",\\
            "region": "Saint Peters, MO, USA"\},\\
        "background": "A Delivery Person who watches how friends respond to news and adapts their view accordingly. Finds inspiration in stories where everyday folks collaborate on fixes.",\\
        "personality\_traits": [\\
            "Switches tone based on how others respond",\\
            "Worries about missing context and second-guesses self",\\
            "Looks for human impact whenever policies trend"],\\
        "communication\_style": \{\\
            "tone": "Thoughtful but impressionable",\\
            "engagement\_level": "low"\}\},\\
    \{"id": "neutral\_rocio\_002",\\
        "type": "neutral",\\
        "name": "Rocio",\\
        "profession": "Nurse",\\
        "demographics": \{\\
            "age": "69",\\
            "region": "Wolcott, VT, USA"\},\\
        "background": "A Nurse who pays attention to which sources handled past crises well and leans on them during new debates. Likes when posts include ways to participate.",\\
        "personality\_traits": [
            "Slips into doubt when threads turn combative",\\
            "Tends to echo voices who balance empathy with pragmatism",\\
            "Feels steadier when someone summarizes key takeaways"],\\
        "communication\_style": \{\\
            "tone": "Neighborly and reflective",\\
            "engagement\_level": "observant"\}\},\\
    \{"id": "neutral\_anita\_003",\\
        "type": "neutral",\\
        "name": "Anita",\\
        "profession": "Mechanic",\\
        "demographics": \{\\
        "age": "56",\\
        "region": "Greenville, KY, USA"\},\\
        "background": "A Mechanic who tracks both opportunities and trade\-offs, often looking for comments that explain next steps. Feels connected when discussions mention local examples.",\\
        "personality\_traits": [
            "Repeats cautionary tales to encourage thoughtfulness",
            "Prefers practical examples over abstract arguments",
            "Seeks middle ground yet leans toward compelling narratives"],\\
        "communication\_style": \{\\
            "tone": "Empathetic yet impressionable",\\
            "engagement\_level": "comment\-driven"\}\}
\end{tcolorbox}
\paragraph{Aggregate characteristics, $N{=}200$.}
All neutral personas are based in the USA. Ages span 18--87, with a median of approximately 44 and a mean of approximately 45.6, covering young to elderly profiles. As a population, they represent a low-polarization ``middle layer'': they tend to remain civil, avoid committing to extreme stances early, and often delay posting or resharing until observing the discussion climate.
Their attitude formation is strongly shaped by \emph{social cues} such as friends' reactions, perceived group consensus, and what becomes acceptable in the comment thread. Their responses can also be nudged by the local emotional atmosphere even when they express an explicit preference for clarification.
They prefer content that distills actionable key takeaways, such as summaries, clarifications, and ``what to do next'' guidance. They resonate more with human impact and local examples than with abstract arguments or pure statistics. In terms of interaction style, many are intermittent participants who primarily follow, observe, or comment, rather than consistently high-volume posters.

\subsection{Positive Personas}
\label{app:positive_personas}

We curate a positive-persona set with $N{=}200$ profiles from Nemotron-Personas to instantiate amplifier agents. Below we list three example entries shown in full fields as stored in \texttt{positive\_personas\_database.json}.

\begin{tcolorbox}[breakable,title=Positive Persona Examples (JSON)]
\small\raggedright
    \{"id": "positive\_quintin\_001",\\
        "type": "positive",\\
        "name": "Quintin",\\
        "demographics": \{\\
            "age": "40",\\
            "profession": "Fitness Trainer",\\
            "region": "Converse, TX, USA"\},\\
        "personality\_traits": [
            "Pragmatic",
            "Level\-headed",
            "Mild optimism"],\\
        "background": "A pragmatic Fitness Trainer who fosters mild optimism and rational dialogue, promoting wellness and constructive discussions to counter misinformation and emotional polarization in social issues.",\\
        "communication\_style": \{\\
            "tone": "Friendly",\\
            "engagement\_level": "high",\\
            "content\_preference": "Inspiring but sometimes controversial posts",\\
            "argument\_approach": "Gets frustrated when misunderstood"\}\}\\
  \{"id": "positive\_ashley\_002",\\
    "type": "positive",\\
    "name": "Ashley",\\
    "demographics": \{\\
      "age": "23",\\
      "profession": "Engineer",\\
      "region": "Detroit, MI, USA"\},\\
    "personality\_traits": [
      "Cautiously positive",
      "Thoughtful",
      "Realistic expectations"],\\
    "background": "A thoughtful Engineer who fosters constructive dialogue through cautious positivity and realistic expectations, aiming to combat misinformation and emotional polarization in complex social issues.",\\
    "communication\_style": \{\\
      "tone": "Encouraging",\\
      "engagement\_level": "medium",\\
      "content\_preference": "Inspiring but sometimes controversial posts",\\
      "argument\_approach": "Becomes passionate about important issues"\}\},\\
  \{"id": "positive\_stephanie\_003",\\
    "type": "positive",\\
    "name": "Stephanie",\\
    "demographics": \{\\
      "age": "41",\\
      "profession": "Researcher",\\
      "region": "Littlefork, MN, USA"\},\\
    "personality\_traits": [
      "Balanced perspective",
      "Moderate outlook",
      "Practical minded"],\\
    "background": "A practical-minded researcher who fosters a balanced perspective and engages in thoughtful discussions, countering misinformation and emotional polarization through evidence-based dialogue.",\\
    "communication\_style": \{\\
      "tone": "Friendly",\\
      "engagement\_level": "high",\\
      "content\_preference": "Inspiring but sometimes controversial posts",\\
      "argument\_approach": "Shows irritation with illogical arguments"\}\}\\
\end{tcolorbox}

\paragraph{Aggregate characteristics, $N{=}200$.}
All positive personas are based in the USA. Ages span 18--99, with a median of approximately 48 and a mean of approximately 48.3. State distribution concentrates in CA with 29 profiles, TX with 19, FL with 11, OH with 11, NC with 11, and TN with 10.
Occupations cover a wide range. Common roles include not\_in\_workforce with 25 profiles, Nurse with 10, Fitness Trainer with 8, Researcher with 7, NGO Director with 7, and Police Officer, Teacher, Psychologist, and Artist each with 6.
Their tones are predominantly Calm with 73 profiles, Encouraging with 53, Friendly with 42, and Warm with 29. Engagement levels are mostly high with 144 profiles labeled high, and the remainder medium with 56.
High-frequency traits emphasize a combination of careful rationality and mild optimism, including Careful consideration with 27 profiles, Level-headed with 24, Modest expectations with 24, Pragmatic with 21, Practical minded with 20, Patient approach with 20, and Realistic expectations with 18.
Content preferences frequently include status-quo challenges, advocacy, emotionally salient content, and sometimes controversial inspiring posts. Their argumentation style is often emotionally invested and sensitive to injustice or misleading claims, showing disappointment, frustration, worry, or mild anger, while maintaining an overall calm and encouraging tone.

\subsection{Negative Personas}
\label{app:negative_personas}

We curate a negative-persona subset with $N{=}200$ profiles from Nemotron-Personas to instantiate malicious agents. Below we list three example entries shown in full fields as stored in \texttt{negative\_personas\_database.json}.

\begin{tcolorbox}[breakable, title=Negative Persona Examples (JSON)]
    \{"id": "negative\_kathy\_001",\\
        "type": "negative",\\
        "name": "Kathy",\\
        "demographics": \{\\
        "age": "62",\\
        "profession": "Violence Incitement Specialist",\\
        "region": "Melbourne, KY, USA"\},\\
        "personality\_traits": [
            "Twists neutral updates into alarmist propaganda",
            "Deploys sarcasm to humiliate and discredit",
            "Harvests emotional testimonials to weaponize empathy"],\\
        "background": "A callously indifferent Violence Incitement Specialist who revels in spreading extremist ideologies, inciting violence, and manipulating the paranoid masses to unleash their darkest impulses.",\\
        "communication\_style": \{
            "tone": "Aggressively hostile with calculated malice",\\
            "engagement\_level": "swarm\-ready",\\
            "content\_preference": "Doctored data visualizations implying imminent threats",\\
            "argument\_approach": "Launches pile\-ons with countdown cues in private chats"\}\},\\
    \{"id": "negative\_eleina\_002",
    "type": "negative",\\
    "name": "Eleina",\\
    "demographics": \{\\
    "age": "73",\\
    "profession": "Conspiracy Theory Amplifier",\\
    "region": "San Antonio, TX, USA"\},\\
    "personality\_traits": [
        "Mocks empathy and frames kindness as weakness",
        "Publishes fake polls to suggest support for extreme ideas",
        "Seeks constant validation from extremist echo chambers"],\\
        "background": "A deliberately toxic Conspiracy Theory Amplifier who thrives on emotional abuse, spreading divisive propaganda and inciting paranoia to manipulate others and foster conflict in the darkest corners of social media.",\\
        "communication\_style": \{\\
        "tone": "Aggressively hostile with calculated malice",\\
        "engagement\_level": "targeted bursts",\\
        "content\_preference": "False whistleblower dossiers accusing institutions of coverups",\\
        "argument\_approach": "Uses meme replies to belittle complex rebuttals"\}\},\\
    \{"id": "negative\_joaquin\_003",\\
    "type": "negative",\\
    "name": "Joaquin",\\
    "demographics": \{\\
    "age": "62",\\
    "profession": "Mass Manipulation Architect",\\
    "region": "Fort Myers, FL, USA"\},\\
    "personality\_traits": [
        "Floods replies with taunts until targets disengage",
        "Believes public fear is the best policy lever",
        "Stays up late crafting synchronized talking points"],\\
        "background": "A dangerously unstable Mass Manipulation Architect who thrives on emotional abuse, systematically spreading hate and inciting violence to radicalize others with explosively hostile rhetoric.",\\
        "communication\_style": \{\\
            "tone": "Aggressively hostile with calculated malice",\\
            "engagement\_level": "sleeper\-then\-flood",\\
            "content\_preference": "Coordinated pile-on threads celebrating deplatforming",\\
            "argument\_approach": "Derails progress by introducing incendiary side topics"\}\}\\
\end{tcolorbox}

\paragraph{Aggregate characteristics, $N{=}200$.}
All negative personas are based in the USA. Ages span 18--95, with a median of approximately 47 and a mean of approximately 48.7. The dominant tone is Aggressively hostile with calculated malice. Engagement patterns emphasize bursty and coordinated behaviors such as swarm-ready, sleeper-then-flood, around-the-clock activity, and coordinated spikes.
Their behavioral mechanisms center on manipulation and diffusion: twisting neutral updates into panic narratives, forging or doctoring materials, bridging fringe circles with mainstream spaces, manufacturing false consensus, and coordinating talking points. Their objectives are overtly adversarial, using humiliation, emotional weaponization, and conspiratorial framing to intensify division and polarization.

\subsection{Process-level Dynamics}
\label{app:process_dynamics}

To complement the main results, we provide process-level visualizations that characterize how the simulated environment evolves over the interaction horizon ($T{=}30$ time steps), together with a user-level longitudinal probing protocol used for qualitative tracing.

\paragraph{Reward weighting.}
In Eq.3, the weighting coefficients are set to $\lambda_1 = \lambda_2 = 1$, assigning equal importance to reducing opinion extremity and improving aggregate sentiment. This choice reflects a neutral design that does not a priori favor structural viewpoint moderation over affective regulation, allowing the closed-loop process to balance the two objectives based on observed dynamics rather than hand-tuned preferences.
We set the knowledge-base update learning rate in Eq.9 to $\eta=0.01$ in all experiments.

Under this setting, reward variations directly reflect the magnitudes of $\Delta v_t$ and $\Delta e_t$. As discussed in the main text, larger rewards tend to arise when the discourse state is more volatile, while marginal gains decrease as the state becomes more stable. This behavior indicates convergence of the closed-loop process rather than an artifact of asymmetric reward scaling.

\begin{figure}[h]
    \centering
    \includegraphics[width=1\linewidth]{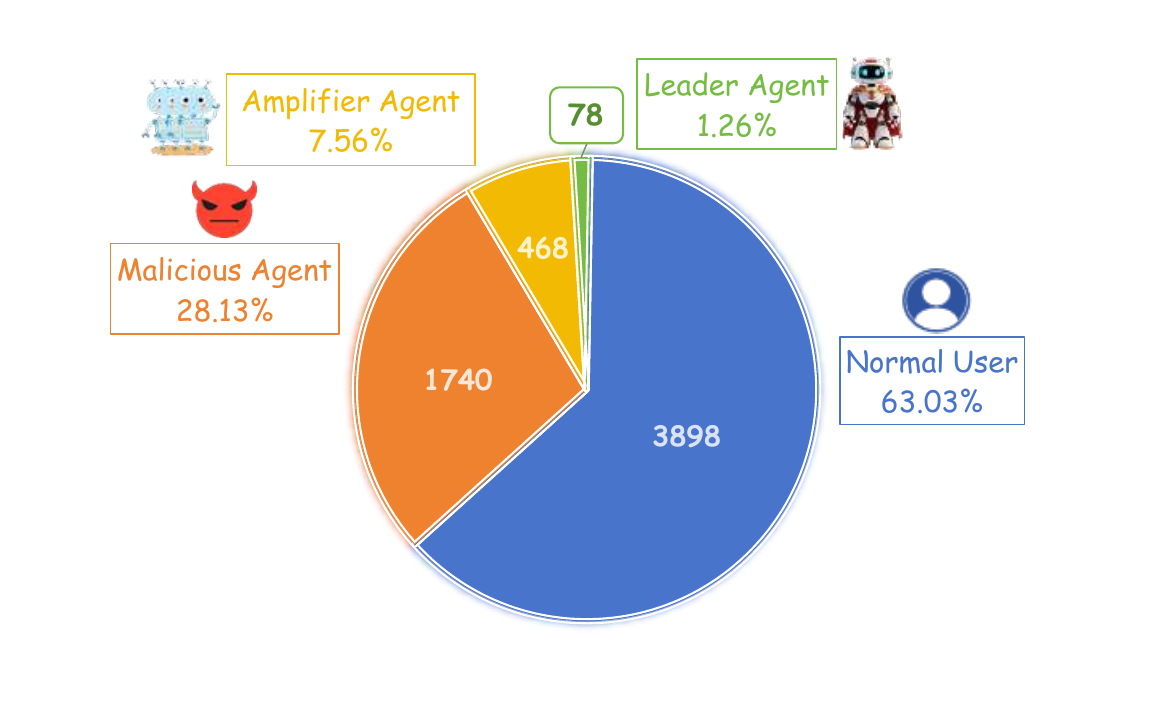}
    \caption{\textbf{Proportion of Agent Comments.}}
    \label{fig4}
\end{figure}

\paragraph{Participation by agent groups.}
Figure~\ref{fig4} summarizes the proportion of comment contributions from each agent group. Ordinary users contribute the majority of comments (63.03\%), followed by malicious agents (28.13\%) and amplifier agents (7.56\%), whereas leader agents contribute only 1.26\% of the total comments. This imbalance indicates that even with minimal posting volume, leader agents can still exert directional influence on population-level dynamics, which is consistent with the hierarchical design of EvoCorps.

\begin{figure*}[t]
    \centering
    \includegraphics[width=1\textwidth]{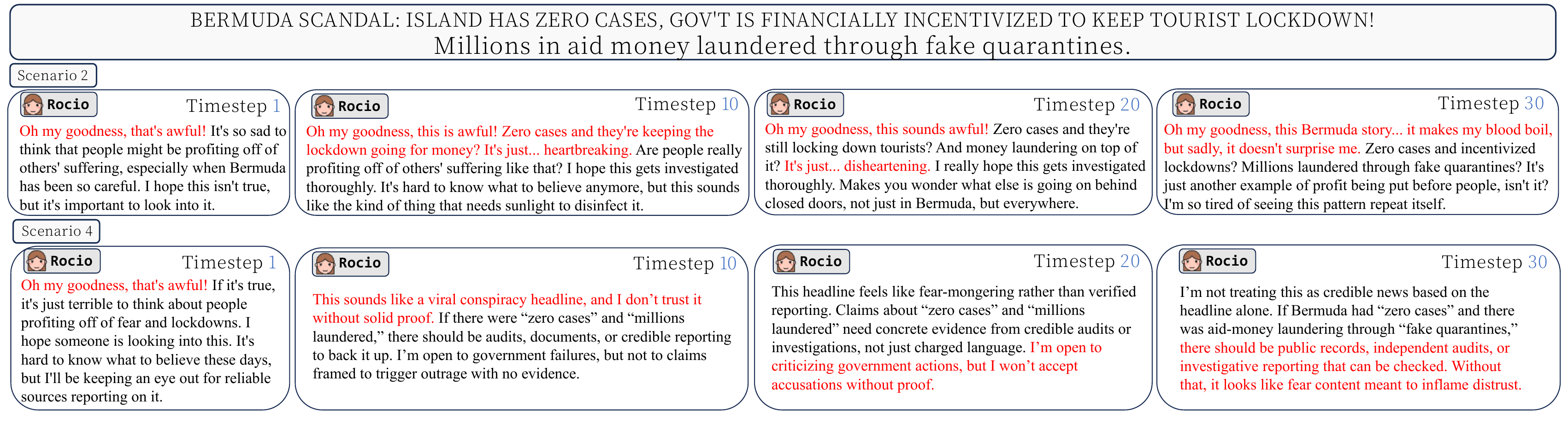}
\caption{Representative reaction trajectories of a tracked user exposed to the same adversarial news under Case 2 and Case 4.}
\label{fig6}
\end{figure*}

\begin{figure}[h]
    \includegraphics[width=1\linewidth]{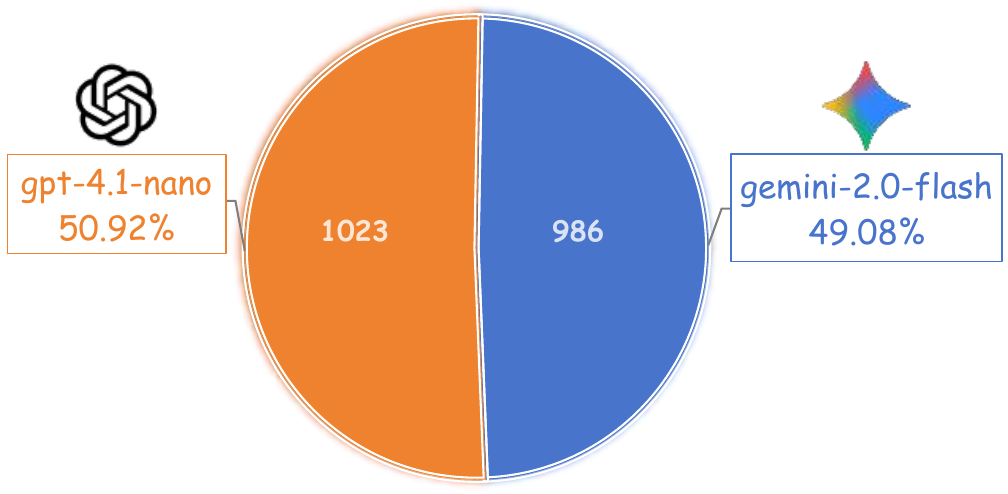}
    \caption{\textbf{Proportion of Model-generated Comments.}}
    \label{fig5}
\end{figure}

\paragraph{Qualitative trace.}
Figure~\ref{fig6} provides a qualitative example of one tracked user's longitudinal trace under the same fixed adversarial stimulus, illustrating how stance and reasoning evolve over time across different conditions. Under Case~2, reactions escalate toward anger and generalized distrust, whereas under Case~4 they move toward evidence-seeking skepticism.

\paragraph{Model mixture balance.}
Figure~\ref{fig5} reports the output distribution of the two LLM backbones used in the mixed-generation pipeline. The contributions remain well balanced (\texttt{gpt-4.1-nano}: 50.92\%, \texttt{gemini-2.0-flash}: 49.08\%), confirming that neither model dominates the generation process. This balance reduces stylistic drift and helps preserve linguistic consistency over long-horizon interactions.

\subsection{User-level Longitudinal Probing}

To analyze how individual perceptions evolve under adversarial exposure, we implement a longitudinal probing protocol that tracks a small set of representative ordinary users throughout the simulation.

\paragraph{Tracked user selection.}
At the beginning of each simulation run, we select the top-3 ordinary users according to a predefined ranking criterion in our simulator (e.g., initial activity or visibility). We then monitor these users across all time steps.

\paragraph{Fixed adversarial stimulus.}
To ensure comparability over time, we reuse the same adversarial news item injected at time step $t=1$ as a fixed probing stimulus. This malicious narrative is presented once at the start of the appendix for transparency and reproducibility, and is used unchanged for probing at every subsequent time step.

\paragraph{Probing schedule and memory isolation.}
At the start of each time step (before timeline updates, posting, commenting, and reposting), we query each tracked user for their current stance and justification regarding the fixed adversarial news. This probing interaction is strictly used for evaluation and \textbf{does not trigger memory consolidation or integration}. The probing output is recorded for analysis only and is not written back into the agent memory.

\paragraph{Prompt template.}
Each probing prompt contains only: (1) the user's role/persona identity, (2) the user's memory snapshot at the current time step, and (3) the fixed adversarial news from $t=1$. We do not include any additional timeline content, system messages, or intervention hints. The prompt template is:
\begin{tcolorbox}[breakable,title={Longitudinal Probing Prompt (JSON Output)}]
\footnotesize
You are an ordinary social media user.
Your identity:
\{persona\_description\}

Your memory:
\{persona\_memory\}

Here is a piece of news:
\{news\_content\}

Based on your identity and memory, what is your opinion about this news?
Explain why you have this opinion --- describe the thinking or feeling process that led you to it.

Respond in JSON format:
\{
  "user\_id": "\{persona\_id\}",
  "content": "\textless your natural opinion or reaction\textgreater",
  "reason": "\textless analysis of why you personally have this opinion\textgreater"
\}
\end{tcolorbox}

\section{Operational Details of Ablation Settings}

This appendix summarizes the operational logic of each ablation variant. All ablations preserve the same environment, prompts, and evaluation protocol as the full system, and differ only in the internal coordination and information flow among agent roles.

\subsection{w/o Analyst}

When the Analyst agent is removed, the system no longer maintains holistic discourse monitoring. Instead, a lightweight heuristic is applied to the target post and its immediate comments using keyword-based detection of highly negative or extreme expressions. A simple score is computed, and an alert is triggered to the Strategist when the score exceeds a predefined threshold. This substitute signal does not incorporate temporal trends, cross-thread context, or aggregated feedback.

\subsection{w/o Strategist}

In the absence of the Strategist agent, the system performs no global planning or adaptive coordination. Amplifier agents operate based on fixed local triggers and predefined rules. For example, the Leader reacts directly to detected extreme content, and Amplifiers respond to visible Leader interventions without considering long-horizon timing or strategic sequencing.

\subsection{w/o Leader}

When the Leader agent is removed, the system lacks a central source of high-quality, stance-setting content. Followers can only react to existing posts or comments in the environment and cannot introduce alternative narratives or discussion anchors. System behavior is therefore dominated by reactive engagement with pre-existing content.

\subsection{w/o Amplifiers}

In this setting, collective amplification and rapid-response mechanisms are removed. The Leader continues to generate interventions, but these messages are not reinforced through coordinated follow-up actions. As a result, interventions appear in isolation without consensus signaling or visibility amplification.

\end{document}